\documentclass[twocolumn]{aastex62}
\usepackage{natbib}
\graphicspath{{./}{../pictures/}}

\received{}
\revised{}
\accepted{}
\submitjournal{ApJ}

\shorttitle{{\sc eccentric disk}}
\shortauthors{Ryu et~al.}
\usepackage{newtxtext,newtxmath}
\usepackage{natbib, amsmath, color}
\usepackage{graphicx, hyperref}

\usepackage[T1]{fontenc}

\DeclareRobustCommand{\VAN}[3]{#2}
\let\VANthebibliography\thebibliography
\def\thebibliography{\DeclareRobustCommand{\VAN}[3]{##3}\VANthebibliography}

\usepackage{graphicx}	
\usepackage{amsmath}	
\usepackage{amssymb}	
\usepackage{xcolor,ulem}
\usepackage{cases}
\usepackage{array,multirow}
\usepackage{upgreek}
\usepackage{textcomp}

\newcommand{\cm}{~{\rm cm}}

\newcommand{\Msol}{~M_{\odot}}

\begin{document}
	
	\title{The Impact of Shocks on the Vertical Structure of Eccentric Disks}
	\correspondingauthor{Taeho Ryu}
	\email{tryu2@jhu.edu}
	
	\author[0000-0002-0786-7307]{Taeho Ryu}
	\affil{Physics and Astronomy Department, Johns Hopkins University, Baltimore, MD 21218, USA}
	
	\author{Julian Krolik}
	\affiliation{Physics and Astronomy Department, Johns Hopkins University, Baltimore, MD 21218, USA}

    \author{Tsvi Piran}
    \affiliation{Racah Institute of Physics, Hebrew University of Jerusalem, Jerusalem 91904, Israel}

\begin{abstract}

Accretion disks whose matter follows eccentric orbits can arise in multiple astrophysical situations.  Unlike circular orbit disks, the vertical gravity in eccentric disks varies around the orbit. In this paper, we investigate some of the dynamical effects of this varying gravity on the vertical structure using $1D$ hydrodynamics simulations of individual gas columns assumed to be mutually non-interacting.  We find that time-dependent gravitational pumping generically creates shocks near pericenter; the energy dissipated in the shocks is taken from the orbital energy. Because the kinetic energy per unit mass in vertical motion near pericenter can be large compared to the net orbital energy, the shocked gas can be heated to nearly the virial temperature, and some of it becomes unbound. These shocks affect larger fractions of the disk mass for larger eccentricity and/or disk aspect ratio. If the orbit can be maintained despite orbital energy loss, diverse initial structures evolve in only a few orbits so that they follow a limit-cycle characterized by a low-entropy midplane and a much higher entropy outer layer. In favorable  cases (such as the tidal disruption of stars by supermassive black holes), these effects could be a potentially important energy dissipation and mass loss mechanism. 
\end{abstract}








\section{Introduction}

Eccentric disks can exist in a variety of astrophysical situations. 
The eccentric rings of Uranus can be observed directly \citep[e.g.,][]{French+1988}. In binary systems, dynamical instability \citep{Lubow1991} caused by tidal perturbation of a companion can lead to eccentricity growth in circumbinary \citep[e.g.,][]{Papaloizou+2001} as well as circum-{single} disks \citep[e.g.,][]{Whitehurst1988}. {Quite a different example can be found} when stars on nearly radial orbits are disrupted by the tidal {gravity} of supermassive black holes; the bound  debris can then form a highly eccentric disk \citep[e.g.,][]{Shiokawa+2015,Sadowski+2016,Svirski+2017,Liu+2017,Cao+2018,Hung+2020,ZanazziOgilvie2020}.

Most previous studies of accretion disks have focused exclusively on circular disks. Moreover, even in the work on eccentric disks, very little attention has been paid to their vertical structure, {even though} it is closely related to observational signatures.
The picture as it stands as of now, largely based on analytic work \citep[e.g.,][]{Ogilvie+2014,Ogilvie+2019,ZanazziOgilvie2020,LynchOgilvie2021}, is that a gas column following an elliptical orbit feels a much stronger vertical gravity at pericenter than at apocenter.  It therefore compresses significantly near the pericenter, by several orders of magnitude for some parameters.  Because $r_{\rm apo}/r_{\rm peri} = (1+e)/(1-e)$ and $g_z \propto r^{-3}$, the near-pericenter compression is greatest when $e$ is comparatively large; it is also enhanced by a more compressible equation of state ($\gamma$ not far from 1). Although it is possible, within the terms of this framework, to find periodic solutions for the time-dependent vertical structure, in these solutions the gas cannot respond quickly enough to reach hydrostatic equilibrium with the local gravity anywhere.

However, most of the previous analytic work \citep[e.g.,][]{Ogilvie+2014,ZanazziOgilvie2020} rests on two strong assumptions: uniform, unchanging entropy; and characterization of the vertical structure by a single parameter, the scale height.   Recently, \citet{LynchOgilvie2021} analytically investigated the dynamic structure of very eccentric disks using a model in which, like the earlier work, the disk's vertical structure evolves homologously, but the entropy can change via heating due to an isotropic $\alpha$-viscosity and radiative cooling. As we show in Section~\ref{subsubsec:overview}, the vertical motions can be expected to reach supersonic speed, raising the prospect that shocks are likely.  If so, both the assumption of constant entropy and the assumption of homologous evolution fail. To explore these possibilities, we have carried out an extensive suite of $1D$ numerical simulations in which we follow a vertical column in its eccentric orbit around a central mass. We find that, as anticipated, shocks do occur in a wide range of the parameter phase space, and once shocks arise, they dramatically alter the evolution of the gas.

These shocks add to the gas's internal energy, resulting in a hotter and more vertically-extended disk. Ultimately, this energy is drawn from the orbit, but in our simulations, its origin lies in the time-dependent potential experienced by the column. In intrinsically thicker disks having larger eccentricity, the energy injected by shocks is so great that the specific energy of the shocked gas can be greater than the instantaneous gravitational potential of the central mass, possibly causing mass-loss from the disk.

We provide a detailed description of our method in Section~\ref{sec:methodology}.
{In Section~\ref{sec:results}, we  describe  our results. Then, we discuss in Section~\ref{sec:discussion} the implications of our results, giving special attention to disks that can form in tidal disruption events (TDEs).
Finally, we conclude with a summary of our findings in Section~\ref{sec:summary}.}

\section{Methodology}\label{sec:methodology}

\subsection{Model}\label{subsec:disk_profile}

We  consider a one dimensional problem of a single column following an orbit of fixed eccentricity $e$ under the gravitational force of a central mass. We assume that this column is isolated in the sense that other material in the disk exerts no forces on it.
The area of the column's footprint does not change, preserving the column's surface density if all its mass remains bound,
as would happen if it were embedded in a disk in which neighboring gas orbits with the same eccentricity and apsidal orientation.

We choose the column's initial location to be at the orbital apocenter, a distance $r_{\rm apo}$ from the central mass.  We assume its initial state is in hydrostatic equilibrium so that
\begin{align}
    \frac{dP}{dz}& = -\Omega^{2}(r_{\rm apo}) z \rho \ .
\end{align}
Here, $\Omega(r)=\sqrt{GM/r^{3}}$ is the angular frequency of a circular orbit at radius $r$; this expression is valid for $|z| \ll r$. These well-defined initial conditions enable us to test whether the disk can maintain hydrostatic equilibrium  along the eccentric orbit. As we show later, the results are generic and independent of this initial condition because time-evolution leads to a robust attractor state.

We further assume that the entropy is the same at all heights in the column, which allows us to use an adiabatic equation of state  $P=K\rho^{\gamma}$ with a constant $K$. Then we have:
\begin{align}
   \rho& = \rho_{\rm c}\left[1 - \frac{z^{2}}{L^{2}}\right]^{1/(\gamma-1)}\label{eq:rho},
\end{align}
where $\rho_{\rm c}$ is the density at $z=0$ and the half-thickness $L$ of the disk is determined by
\begin{equation}\label{eq:L}
    L =\left[\frac{2\gamma K\rho_{\rm c}^{\gamma-1}}
    {(\gamma-1)\Omega^{2}}\right]^{1/2}.
\end{equation}
Not surprisingly, $L \sim c_s(z=0)/\Omega$. A commonly-used measure of disk vertical thickness is the density-weighted moment $h$, 
\begin{align}\label{eq:scaleheight}
    h&=\frac{\int_{0}^{L}\rho |z| dz}{\int_{0}^{L}\rho dz}.
\end{align}
For disks with adiabatic index $\gamma=4/3$ in hydrostatic equilibrium (described by Equation~\ref{eq:rho}), the height defined by the density moment is $h_{\rm HSE}=0.27L$,  and the surface density is $\Sigma=0.9 ~L ~\rho_{\rm c}$. It follows that the aspect ratio is
\begin{align}\label{eq:aspectratio}
   \frac{h_{\rm HSE}}{r} 
    &=A(\gamma)\left[ \frac{K \Sigma^{\gamma-1}}{r^{\gamma+1}\Omega^{2}}\right]^{1/(\gamma+1)},
\end{align}
where $A(\gamma)$\footnote{$A(\gamma)=\left[\frac{2}{\sqrt{\uppi}^{\gamma}}\left(\frac{\gamma-1}{\gamma}\right)^{\gamma}\frac{~\Gamma[(3\gamma-1)/(2\gamma-2)]^{2}}{\Gamma[\gamma/(\gamma-1)]^{2}}\right]^{1/(\gamma+1)}$}is $\simeq0.78$ for $\gamma=4/3$.

 In all our simulations, the evolving pressure in the column is found from the internal energy through the relation $P=(\gamma-1)\rho e_{\rm int}$
with $\gamma=4/3$, appropriate to a radiation-dominated state.
To test our conclusions for sensitivity to $\gamma$, we also made a few runs with $\gamma=5/3$ (appropriate to a state whose pressure is due to an ideal gas with no internal degrees of freedom).  Compared at the same aspect ratio of the disk, we find no significant sensitivity to $\gamma$.

\begin{figure*}
\centering
		\includegraphics[width=18.0cm]{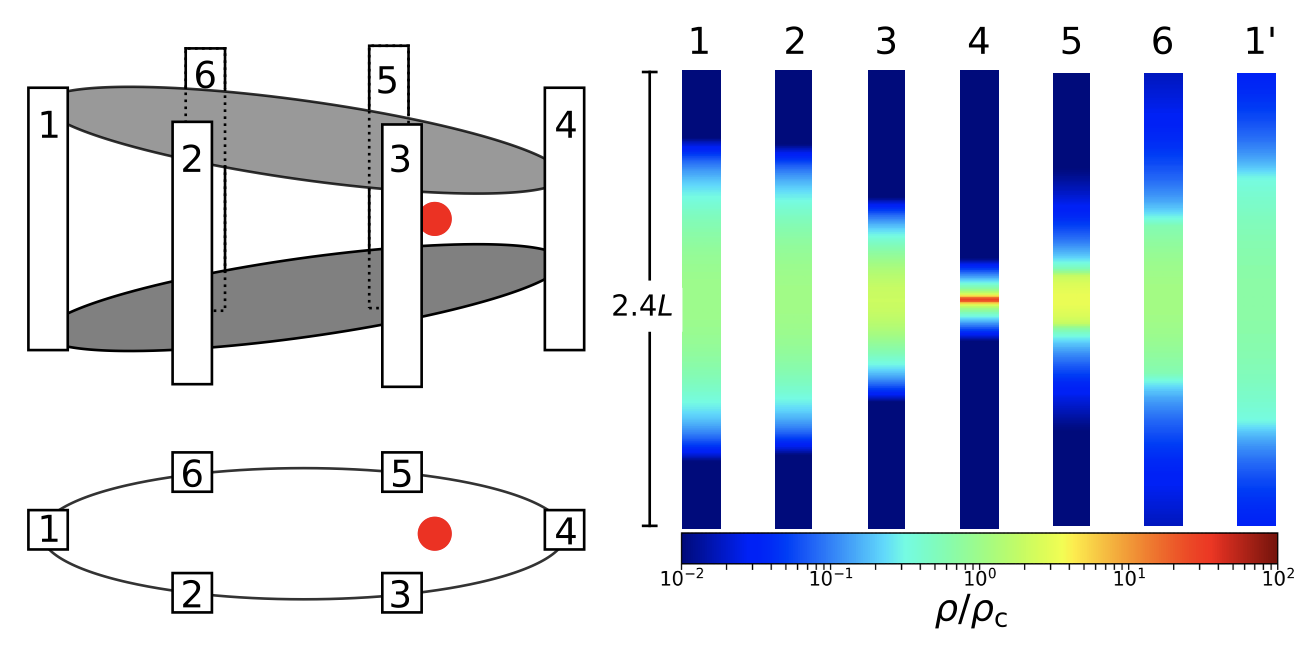}	
		\caption{Schematic diagrams for an elliptical disk around a central mass (red dot) in an edge-on view (\textit{left-top}) and a face-on view (\textit{left-bottom}). We evolve the gas in the column as it orbits (from 1 to 6). On the right, we show the density distribution within $|z|\leq 1.2L$ for the model \textbf{H01} with $e=0.7$ at the first pericenter passage. The label 1' corresponds to the column when it returns to the apocenter after one orbit. }
	\label{fig:overview}
\end{figure*}

\subsection{Orbit and disk parameters}\label{subsubsec:parameters}

Because Newtonian gravity has no special length scale and we have simplified the geometry of this problem to 1D, it can be completely specified by two dimensionless parameters: the initial aspect ratio $h_0/r$, which is determined by its initial heat content; and the orbital eccentricity $e$.  This reduction can be achieved by setting the unit of length to the semimajor axis $a$, the unit of time to the orbital dynamical time $\Omega(a)^{-1}$, and the unit of mass density to the initial central density.

 To explore how such an orbiting gas column's behavior depends on $h_0/r$ and $e$, we consider $3-6$ different eccentricities within the range $0.4\leq e\leq0.9$ and three different initial hydrostatic-equilibrium aspect ratios: $h_0/r\simeq 0.1$ (model \textbf{H01}), $0.03$ (model \textbf{H003}), and $0.001$ (model \textbf{H001}). The parameters of the disks considered in this study are summarized in Table~\ref{tab:parameter}. 
Note that for fixed $a$,
the radius at apocenter depends on eccentricity, making the exact value of $h_0/r$ vary slightly with $e$ within a given model.

We present an overview of the situation in Figure~\ref{fig:overview}.  In the left panel, there are schematic diagrams for both a face-on view and an edge-on view of the {orbiting} column. The numbers 1-6 show successive positions of the column; $1^\prime$ corresponds to a return to apocenter. In the right panel, we show the actual density distribution for our fiducial disk model, \textbf{H01} with $e=0.7$ as the gas column moves around its first complete orbit, having begun with a hydrostatic equilibrium configuration.  It is strongly compressed near pericenter, where a strong shock is visible.  By the time the column returns to apocenter, the column's density distribution has expanded noticeably, particularly in its outer layers.

\begin{table}
	\centering
\begin{center} 
	\renewcommand{\thetable}{\arabic{table}}
	\caption{Disk parameters. For fixed $\rho_{\rm c}$, $K$ and $a$, $h_0/r_{\rm apo}\propto (1+e)^{2/7}$ for $\gamma=4/3$ (Equation~\ref{eq:aspectratio}). Here, $h_{0}$ refers to the initial hydrostatic-equilibrium scale height (Equation~\ref{eq:scaleheight}) and $r_{\rm apo}$ is the apocenter distance. }
	\label{tab:parameter}
	\begin{tabular}{c c c }
	\hline
	Model name &  $h_0/r$ & $e$\\
	\hline
    \textbf{H01}  &	 $0.1-0.11$ & 0.4, 0.5, 0.6, 0.7, 0.8, 0.9\\
    \textbf{H003}  & $0.032-0.036$ & 0.5, 0.7, 0.9\\
    \textbf{H001} &	 $0.01-0.011$ & 0.5, 0.7, 0.9\\
	\hline
	\end{tabular}
	\end{center}
\end{table}

\subsection{Code}\label{subsec:code}

We solve the 1D hydrodynamics equations in a Cartesian coordinate system using the code {\small CASTRO} \citep{Almgren+2010}. {\small CASTRO} is an adaptive mesh, compressible radiation/magneto-hydrodynamics Eulerian code. {\small CASTRO} has its own shock capturing algorithm, built upon the shock condition of \citet{ColellaWoodward1984}. It has been used for solving a variety of astrophysical problems, including mergers of white dwarfs \citep{Katz+2016}, the structures of supernova shocks \citep{Burrows+2012} and exoplanet atmospheres \citep{Ryu+2018}. We do not use its radiation and MHD capabilities in this work. 

The equations we solve in this 1D application of {\small CASTRO} are (before non-dimensionalizing):
\begin{align}
\frac{\partial\rho}{\partial t}&=-\frac{\partial}{\partial z}(\rho u^{z}),\\
\label{eq:hydro2}
\frac{\partial(\rho u^{z})}{\partial t}&=~-\frac{\partial}{\partial z}\left[\rho (u^{z})^{2} +  P\right]+\rho g^{z},\\
\label{eq:hydro3}
\frac{\partial(\rho e_{\rm tot})}{\partial t}&=-\frac{\partial}{\partial z}[\rho u^{z}e_{\rm tot}+Pu^{z}]+\rho u^{z} g^{z},
\end{align}
where $\rho$, $u^{z}$, and $P$ are the density, velocity in the vertical direction, and pressure, respectively. The total specific energy $e_{\rm tot}$ used by the code is the sum of the internal energy $e_{\rm int}$ and the kinetic energy $e_{\rm kin}$ ($=(u^{z})^{2}/2$), i.e., $e_{\rm tot}=e_{\rm int}+e_{\rm kin}$.  Work done by the vertical gravity $g^z$ of the central mass enters through the source term $\rho u^z g^z$.  The vertical gravity $g^z$ is
\begin{align}\label{eq:grav}
   g^{z} =  -\frac{GMz}{(r^{2} + z^{2})^{3/2}},
\end{align}
and the corresponding potential is
\begin{align}
e_{\rm pot}=-\frac{GM}{\sqrt{r^{2}+z^{2}}}+\frac{GM}{r},
\end{align}
 so that $e_{\rm pot}= 0$ at $z=0$.
Here, $r$ is the distance from the central mass $M$ to the center of the column (at $z=0$).

\subsection{Simulation setup}\label{subsec:sim_setup}

Our domain extends from the midplane to $z=4L$, where $L$ is the initial (half)-thickness of the disk (see Section~\ref{subsec:disk_profile}); in 1D, up/down reflection symmetry is maintained automatically.  At the top of the box, we  place an outflow boundary condition
(i.e., zero-gradient extrapolation with no inflow allowed).

Outside the finite extent of the disk gas, we fill the remainder of the simulation box with very low density ($\rho = 10^{-15}\rho_{\rm c}$) and low pressure ($P = 10^{-8}P_{\rm c}$) ``numerical vacuum" material. Occasional unphysical surges in velocity can occur in the numerical vacuum near the boundary
of the domain.  To curb these events, we damp the vertical velocity of all gas with $\rho < 10^{-7}\rho_{\rm c}$ using a ‘sponge’ damping method employed in the low Mach number code {\sc MAESTRO} \citep{Nonaka+2010} and other studies \citep[e.g.,][]{Zingale+2009,Ryu+2018}.
See \citealt{Almgren+2008} for the equations used in this the damping scheme.

With this setup, we find that our disk models  remain in excellent hydrostatic equilibrium (just as described in Section~\ref{subsec:disk_profile}) for more than 20 orbits when $e=0$.   The relative error in density is less than $\sim 10^{-7}$.

The number of cells in our grid increases with $e$ because the gas is more compressed in a more eccentric disk ($10^{4}$ cells for $e\leq 0.5$, $5\times10^{4}$ for $e=0.6$ and $2\times10^{5}$ for $e\geq0.7$ in our standard model \textbf{H01}).  However, the number of cells for $e\geq0.7$ in the thinner disks must be even larger because gas is more compressed in thinner disks ($3-6\times10^{5}$ cells for $e\geq0.7$ in models \textbf{H003} and \textbf{H001}). We have performed convergence tests with resolution coarser and finer by a factor of $1.5-2$ and find no significant differences between those simulations and runs with our standard resolution (see Appendix \ref{appendix:res} for more details about the convergence tests). We consistently use a Courant number of 0.5.

For our fiducial models (\textbf{H01}), we analyze disk evolution for ten complete orbits, which is sufficient for the system to reach a limit cycle. 
In order to understand the dependence on initial thickness, we compare the fiducial models with two thinner disk cases (\textbf{H003} and \textbf{H001}) for the first 
orbit.

\section{Results}\label{sec:results}

\subsection{Overview of dynamical picture}\label{subsubsec:overview}

Before discussing our numerical results, we will begin with an overview of vertical dynamics in eccentric disks.  As the gas column approaches the pericenter, it is compressed by the increasing vertical gravity (see Figure~\ref{fig:overview}).  As shown in Equation~\ref{eq:aspectratio}, if the fluid evolves adiabatically, its hydrostatic scale height, and therefore its equilibrium length $L$, depend on two constant parameters, $K$, and $\Sigma$, so that their variation is solely through the strength of vertical gravity, which is expressed by $\Omega^2 \propto r^{-3}$.
As already noted, $L \sim c_s(z=0)/\Omega$, so in order
to maintain hydrostatic equilibrium, $L$ must change at a rate $\dot L \sim (c_{\rm s,z=0}/\Omega) (\dot r/r)$.
The Mach number of this motion (relative to the $z=0$ sound speed) is therefore $\dot{L}/c_{\rm s,z=0}\sim (\dot r/r)\Omega^{-1}$.
More precisely,
\begin{align}\label{eq:hydrofall}
\frac{|\dot L|}{c_{\rm s,z=0}} =&\frac{4\gamma^{1/2}}{(\gamma-1)^{1/2} (\gamma +1)} \frac{\dot r/r}{\Omega(r)} \nonumber \\ =&\frac{4\gamma^{1/2}}{(\gamma-1)^{1/2} (\gamma +1)}\frac{e|\sin\phi|}{1+e\cos\phi}. 
\end{align}

This characteristic Mach number depends on only the eccentricity, orbital phase, and adiabatic index; it is independent of the disk thickness and the orbital velocity.
The greatest Mach number is therefore found where $r=a$, and its maximal value is $\propto e(1-e^2)^{-1/2}$. Consequently, the compression speed needed to keep the disk in hydrostatic equilibrium becomes supersonic once the eccentricity is moderately large: for $\gamma = 4/3$, the estimate of Equation~\ref{eq:hydrofall} implies supersonic motion when $e \gtrsim 0.28$. 
 Thus, even for modest eccentricity, supersonic motion can be expected.

\begin{figure}
\centering
\includegraphics[width=8.6cm]{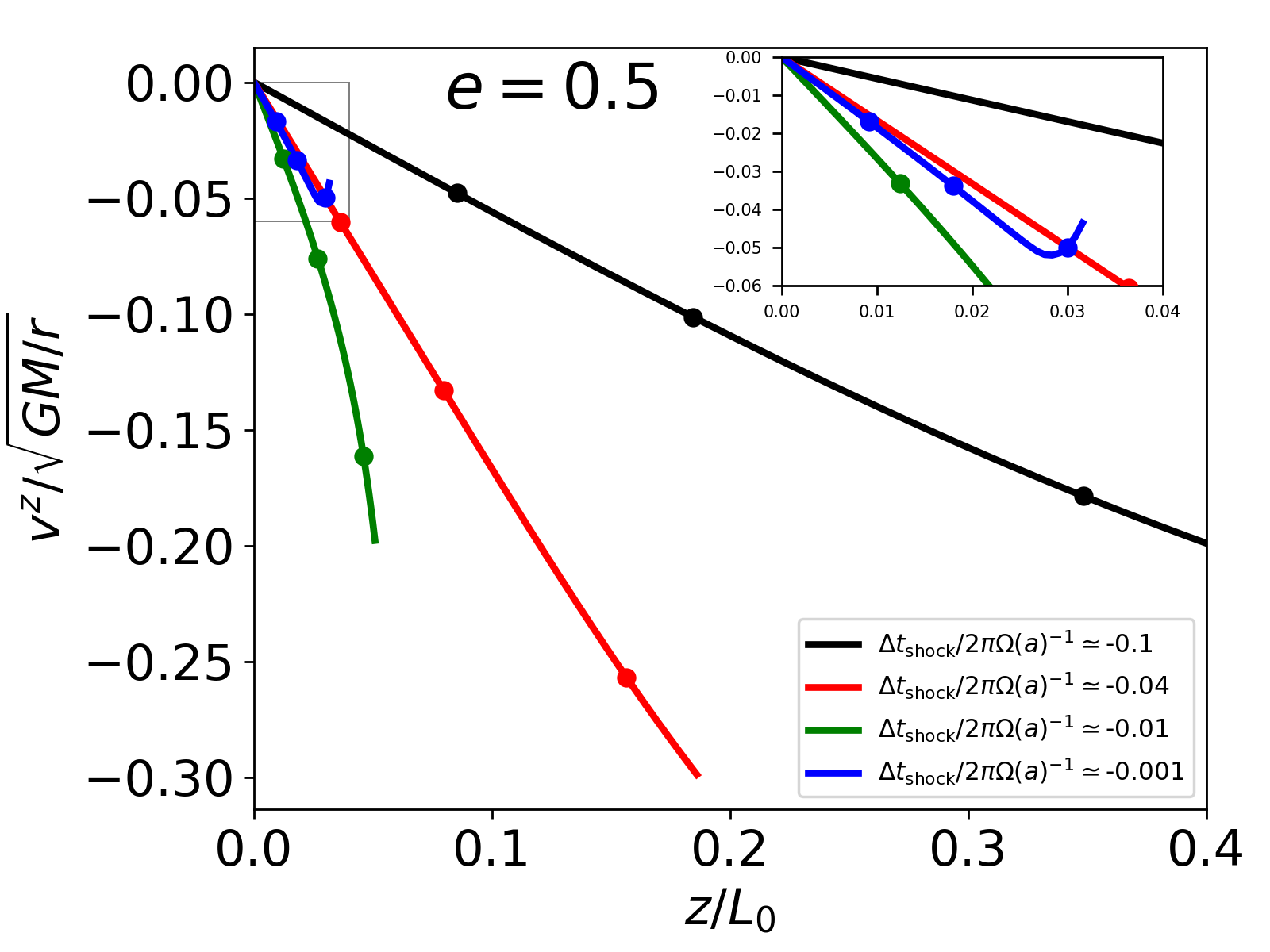}
\includegraphics[width=8.6cm]{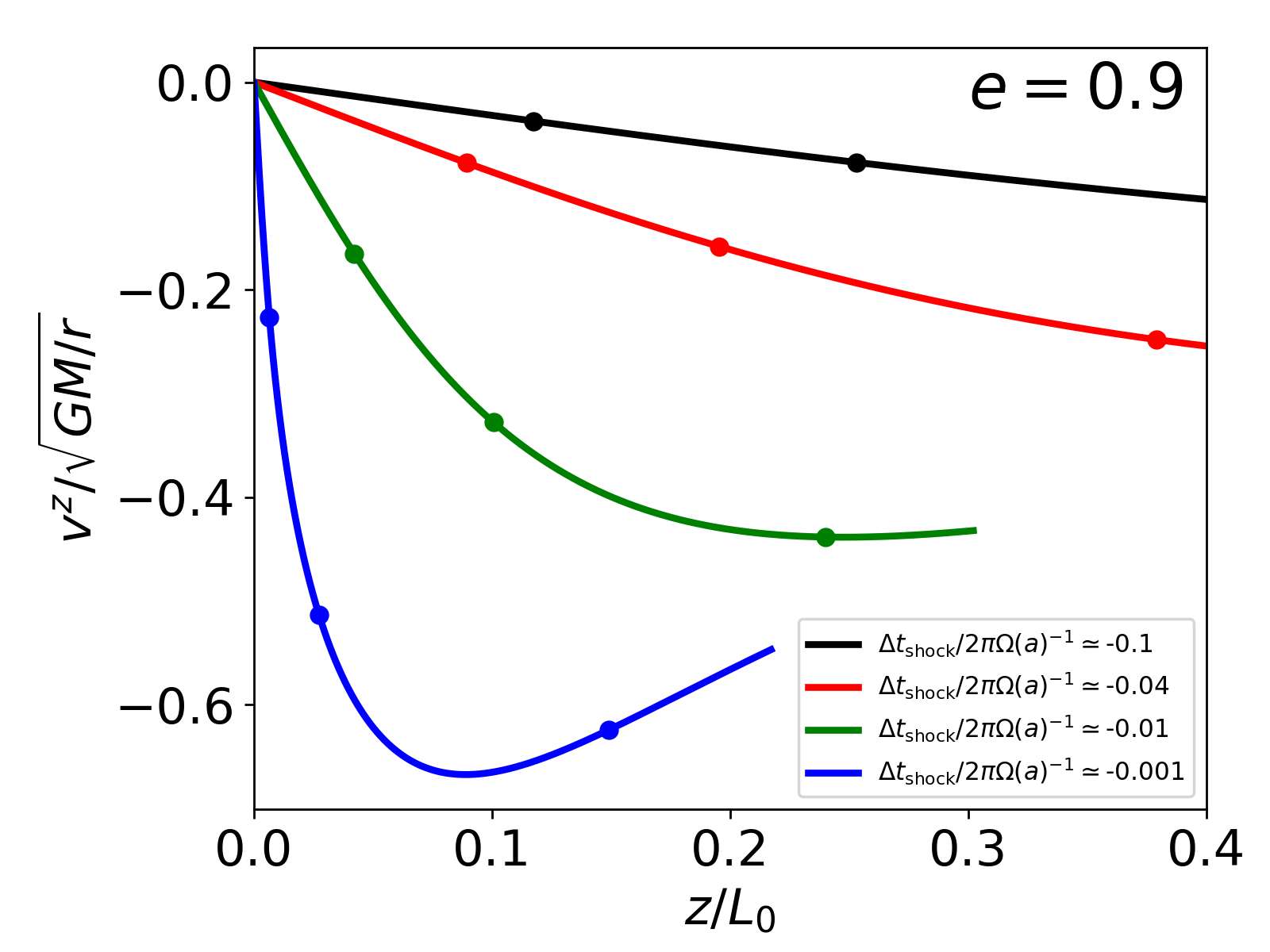}
\caption{The ratio of the vertical velocity to the instantaneous free-fall speed  for $e=0.5$ (\textit{top} panel) and $0.9$ (\textit{bottom} panel) in the model \textbf{H01}, as a function of $z/L_{0}$ at a few different times before the first shock forms. $\Delta t_{\rm shock}$ indicates the time relative to shock formation (negative means before the shock). The circles mark the fractional enclosed mass relative to the initial total column mass, $30\%$, $60\%$, and $90\%$; the curves end at mass-fraction $95\%$. The turn-up at $\Delta t_{\rm shock}/2\uppi\Omega(a)^{-1}$ (blue) for $e=0.5$ can be seen more clearly in the inset.}
	\label{fig:vertical_velocity}
\end{figure}

As our simulations show (see Section~\ref{subsec:vertical}), when the column first starts to evolve, it does so homologously, i.e., maintaining a fixed shape for the vertical profile of all quantities, but with the lengthscale set by a time-dependent scale height.  However, as outer fluid elements fall toward the midplane, they encounter slower-moving  matter; the associated compression creates a pressure gradient, and this decelerates some of the fluid elements.  This deceleration is the origin of the breakdown of homologous behavior. As suggested by Equation~\ref{eq:hydrofall}, it happens sooner and to a larger degree for higher $e$ (Section~\ref{subsec:adiabatic}), but significant departures from homologous behavior begin only very close to the time at which the shock forms because that is the time at which the pressure contrast across distances small compared to the scale height first becomes substantial.   When the pressure gradients steepen sufficiently, they become shocks.  Although the shocks initially propagate from the surface of the disk toward the midplane, after a short time they are reflected and move outward toward the disk surface.  The dissipation associated with these shocks invalidates the assumption of constant entropy.

The kinetic energy of the matter moving vertically within the column is taken from the potential energy due to its initial height above the orbital plane.   Put another way, it is the result of work done by the tidal gravity within the column. 
The shock converts the majority of the gas's kinetic energy to internal energy. In addition, when gas near the middle of the column is compressed toward the midplane, the potential energy it gives up is transferred to gas farther out.  At apocenter, that matter is then able to reach a greater height from the midplane than it did an orbital period earlier.
Although the upward motion of the gas decelerates as it approaches apocenter, some of it retains so much kinetic energy that it becomes unbound (see Section~\ref{sec:expelled}).  

Given the estimate of Equation~\ref{eq:hydrofall}, the impacts of the shocks---heating, energy transfer, and mass expulsion---are greater in more eccentric disks. Similarly, because $g_{\rm z} \propto z$, they are also greater for thicker disks.

Remarkably, all of our disk models reach a stable limit-cycle in which the column alternates between an extended state at apocenter and a highly compressed one at pericenter.  In this state, although strong shocks create large entropy contrasts, and the column steadily loses mass, there is almost no orbit-averaged change in {\it specific} energy.
As will be shown in detail below, the long-term stability of the limit-cycle arises from a balance between shock heating and the energy loss associated with expelling mass.

In the following sections, we  quantitatively elaborate on the remarks in this section.

\begin{figure}
\centering
\includegraphics[width=8.6cm]{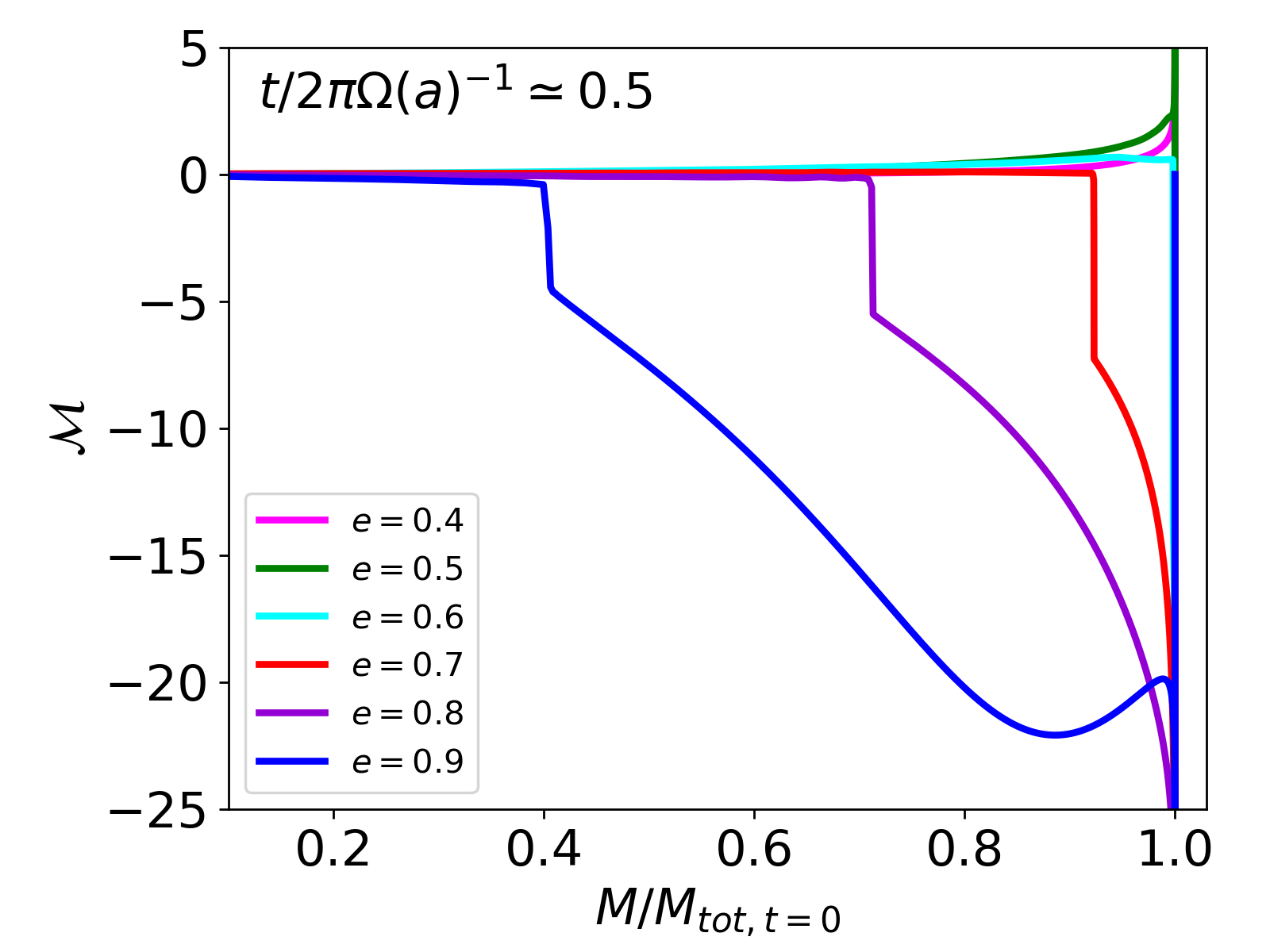}
	\caption{Mach number $\mathcal{M}$ of the gas when the column is most compressed near the first pericenter passage, as a function of the enclosed mass relative to the initial total mass of the column. Negative (positive) values indicate inward (outward) motion.
	}
	\label{fig:Mach_shock}
\end{figure}

\begin{figure}
\centering
	\includegraphics[width=8.6cm]{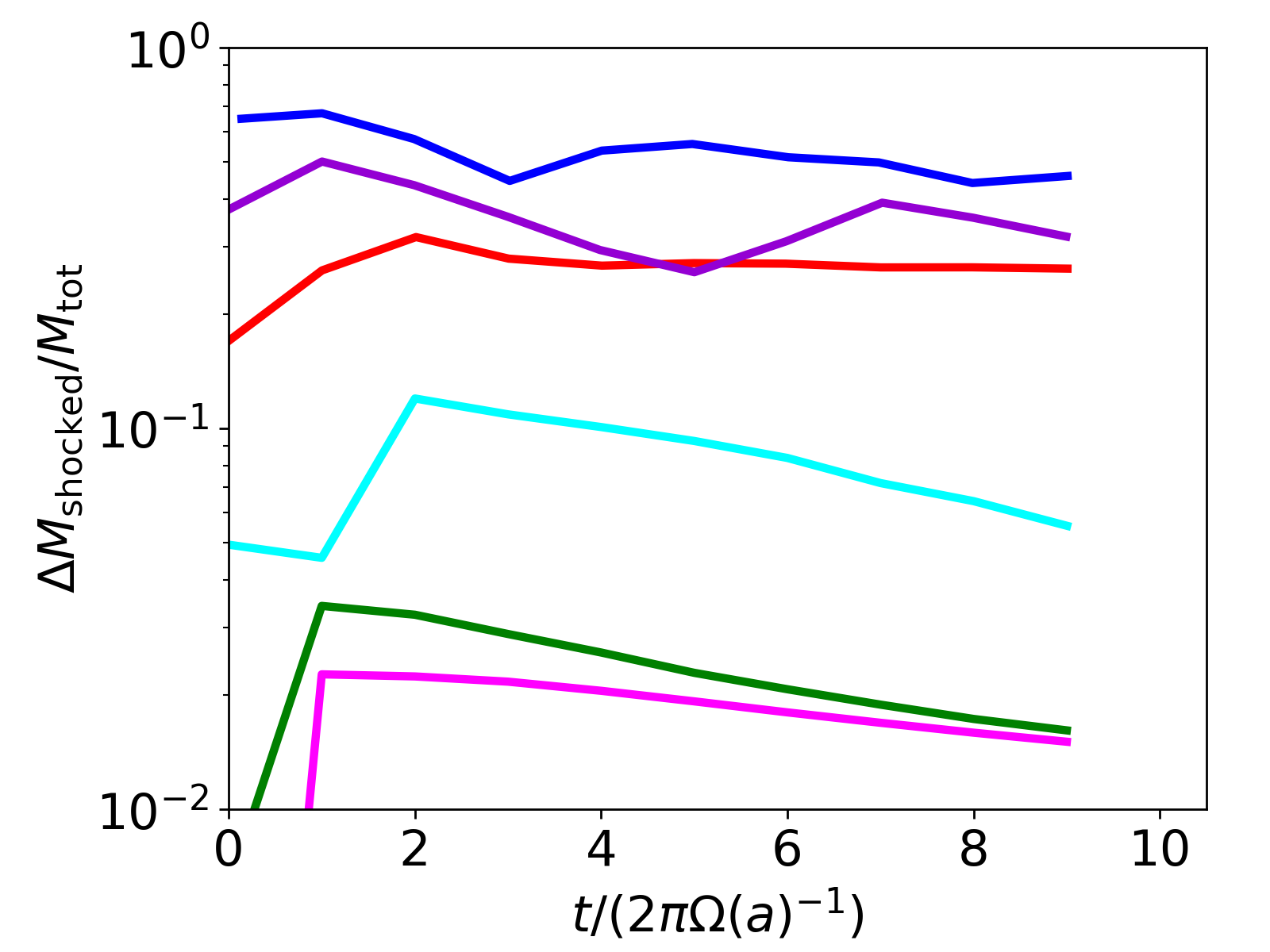}		
    \includegraphics[width=8.6cm]{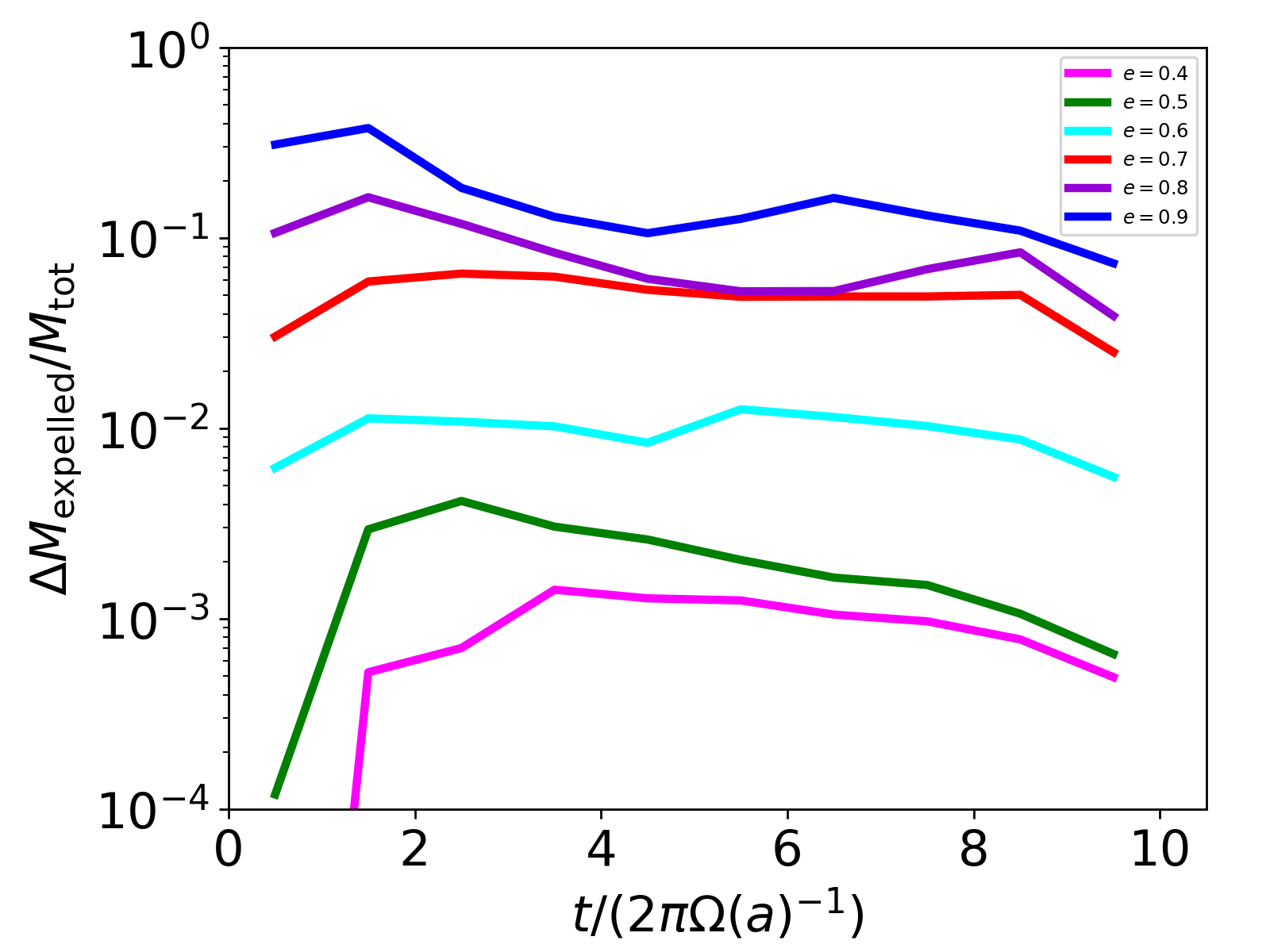}
		\caption{\textit{top} panel : Ratio between the mass of gas shocked in  a given orbit and the total column mass at the beginning of the orbit. \textit{bottom} panel : Ratio between the mass expelled in a given orbit and the column mass at the beginning of that orbit as a function of orbit number. $\Delta M_{\rm expelled}$ is calculated as the difference in the total mass at two adjacent pericenters. Note that the dynamic range in the bottom panel is twice as large as in the top panel. In both panels, $M_{\rm tot}$ is defined as $2\int_{0}^{z_{\rm top}}\rho dz dA$ where $z_{\rm top}$ is the height at the domain outer boundary and $dA=dxdy$.}
	\label{fig:shocked_gas_mass}
\end{figure}

\subsection{Adiabatic evolution}\label{subsec:adiabatic}

Although our simulations begin in hydrostatic equilibrium, the gas column swiftly moves away from equilibrium as it leaves its initial position at apocenter. Initially, the column evolution is both adiabatic and homologous.  Because the logarithmic Lagrangian rate of density change in a fluid element $d\ln\rho/dt = -\partial v/\partial z$, a linear velocity profile leads to homologous behavior; a greater slope indicates more rapid density evolution.  As shown in Figure~\ref{fig:vertical_velocity}, this picture obtains until a short time before the shock forms.  However, the converse also holds: deviation from a linear velocity profile signals that $d\ln\rho/dt$ is a function of position, i.e., the behavior is no longer strictly homologous.  This happens at a time before the shock roughly $10\times$ greater (in fraction of an orbital period) for $e=0.9$ as compared to $e=0.5$.  The deviation is greatest for the mass farthest from the midplane because the initial density profile is steepest there; for isentropic adiabatic gas, greater gas pressure deceleration follows.

\subsection{Shock formation and penetration}\label{subsec:vertical}

The adiabatic stage 
ends when the column nears pericenter and infalling gas collides with more slowly-moving gas at lower altitudes, thereby creating shocks. The shocks are always initiated close to pericenter, but whether it is immediately before or after depends on parameters.  For example, as shown in Figure~\ref{fig:Mach_shock}, which illustrates the Mach number of the gas at the moment of greatest compression, the shocks in the high-eccentricity cases of \textbf{H01} have ${\cal M} \simeq 5 - 8$, and, at this point, are traveling inward, while matter at higher altitude is falling at much higher Mach number.  On the other hand, the shocks in the low-eccentricity cases are much weaker, and at this point have already turned around and reached the outer surface of the column.

These shocks last for only a brief time near the moment of pericenter.  Their durations are so limited because  the shocks are very fast: they move at close to the local circular orbit speed $r_{\rm peri}\Omega(r_{\rm peri})$, accelerating from $\simeq 0.2$ to $\simeq 0.6-0.7\times$ this speed as they rise. However, because the shock passes through a much greater portion of the column when $e \gtrsim 0.6$ than when $e$ is smaller, the duration of the shock passage is a similarly strong function of eccentricity, increasing by a factor $\sim 100$ from $e=0.5$ to $e=0.9$, while still remaining much smaller than an orbital period.
Changing vertical gravity consequently has little effect on the column during the shock's passage.

\begin{figure}
		\includegraphics[width=8.6cm]{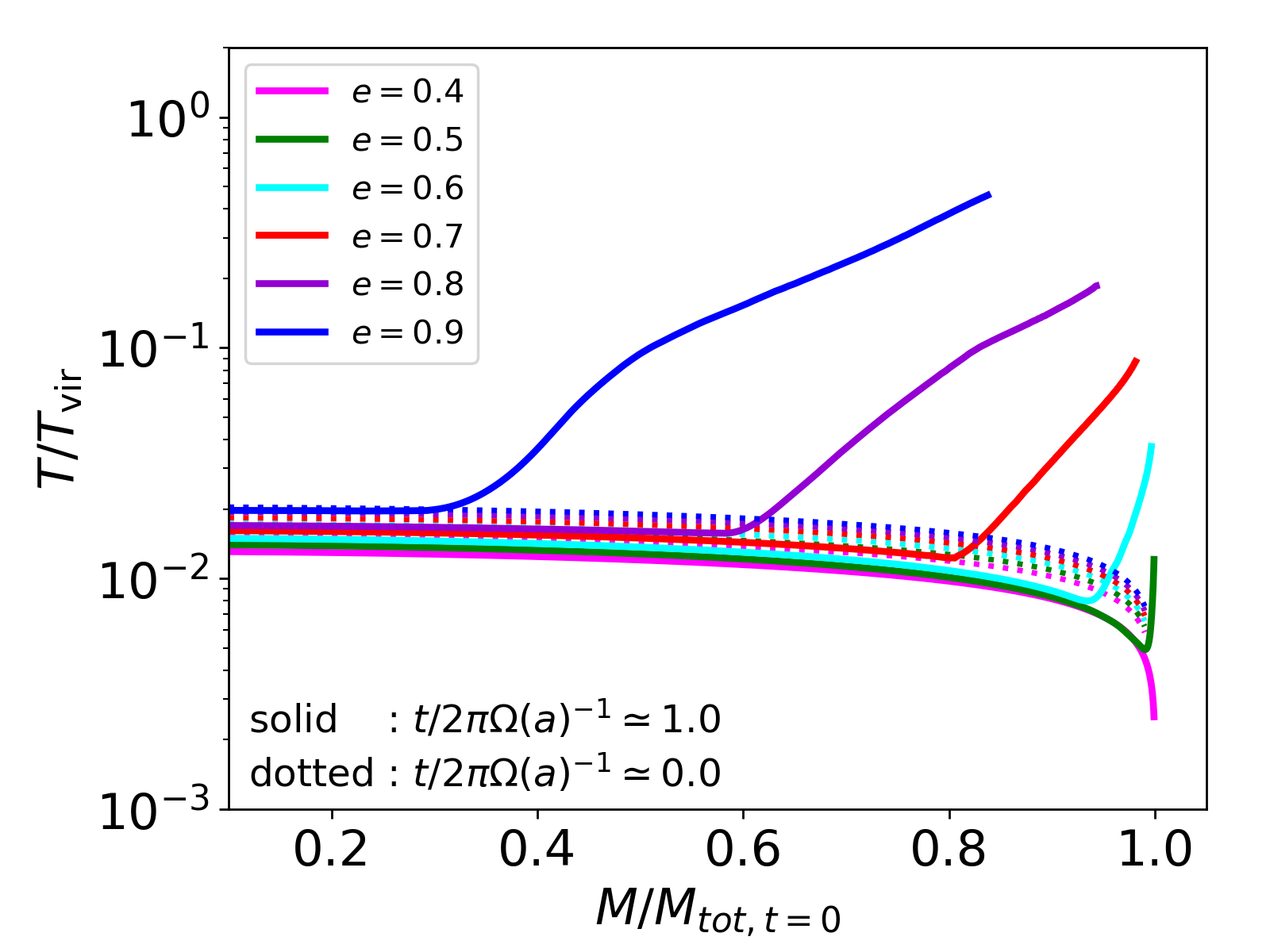}	
		\caption{The temperature $T$ normalized to the apocenter virial temperature  $T_{\rm vir} \equiv GM\mu m_p/[a(1+e)k_B]$, as a function of the enclosed mass relative to the initial total mass.  Six different eccentricities in the model \textbf{H01} are shown at both the initial state and the first apocenter passage. Note that the sharp increase in $T/T_{\rm vir}$ at $M/M_{{\rm tot},t=0}\simeq 1$ for $e\leq0.5$ is due to the shock heating of a small amount of mass near the disk's surface.}
	\label{fig:T_S}
\end{figure}

The eccentricity-dependence of the mass of shocked gas residing in the column is shown in the \textit{upper} panel of Figure~\ref{fig:shocked_gas_mass}.\footnote{We define the shocked mass as those parcels with $K/K_{t=0}>3$ where $K=p/\rho^{\gamma}$.
This is a somewhat arbitrary definition for the shocked material. However, this simple criterion well-identifies the inflection point in $K$ shown in the \textit{left} panel of Figure~\ref{fig:T_S} and is insensitive to the choice of the lower limit of $K/K_{t=0}$.}
Because the deepest point shocked is always within gas that had never previously been shocked, and the shock always runs out from there to the top of the column, the total mass of gas shocked in an orbit is the sum of the previously-shocked gas at the beginning of the orbit and the decrease in never-shocked gas during the orbit.
For $e \lesssim 0.5$, the fraction of a column's mass shocked per orbit is a few percent, more or less independent of time after the first orbit; for $e \gtrsim 0.7$, it is a few tens of percent, rising with eccentricity almost as a step-function.  The only intermediate case is $e=0.6$. 
 After the first orbit, much of the shocked gas has been shocked before, so the mass of never-shocked gas, which is always the gas closest to the midplane, declines slowly, particularly for lower eccentricity.  By contrast, at high eccentricity only a few orbits are required in order to shock the majority of the original gas mass: for $e=0.9$, for example, the unshocked gas is reduced to $27\%$ of the original mass after a single orbit.

\subsection{Shock heating and the column's internal structure}\label{subsec:shockheating}

To illustrate the impact of shock heating, we show in Figure~\ref{fig:T_S} the temperature as a function of the enclosed mass at two times, the initial state and the first return to apocenter, for all the eccentricity cases of model \textbf{H01}.
For the present purpose, we define the gas temperature as $(1/3)e_{\rm int}\mu m_{\rm P}/k_{\rm B}$ because this is the most appropriate criterion for considering the ratio of internal energy to gravitational potential energy when $\gamma = 4/3$; we define the virial temperature $T_{\rm vir}$ as the energy equivalent to the potential energy at the apocenter, measured at $z=0$.  The  ratio $T/T_{\rm vir}$ of the unshocked gas is $\sim 10^{-2}$, but shocked gas (gas above the inflection point) is 1 - 2 orders of magnitude hotter.

\begin{figure}
\includegraphics[width=8.6cm]{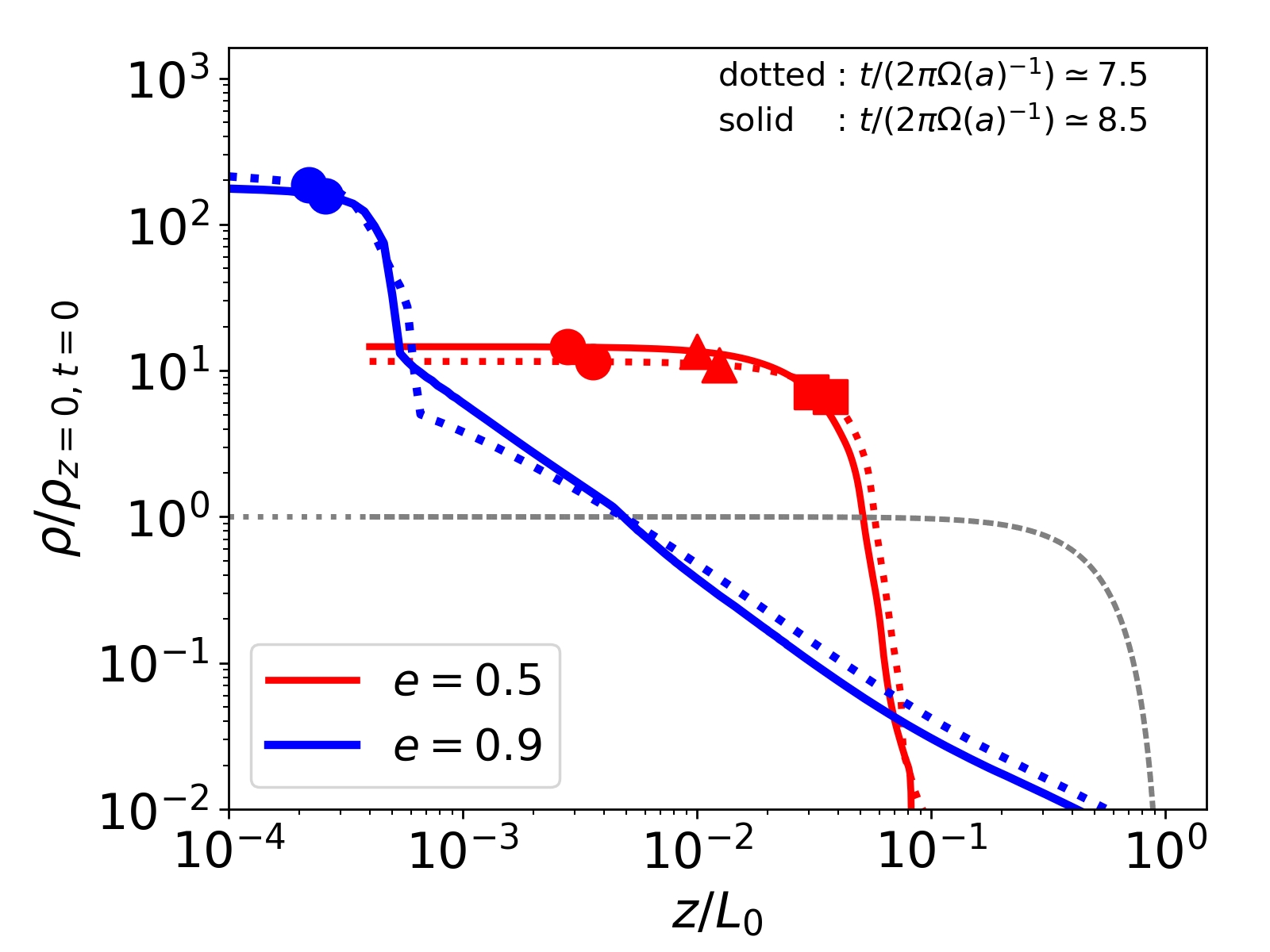}
\includegraphics[width=8.6cm]{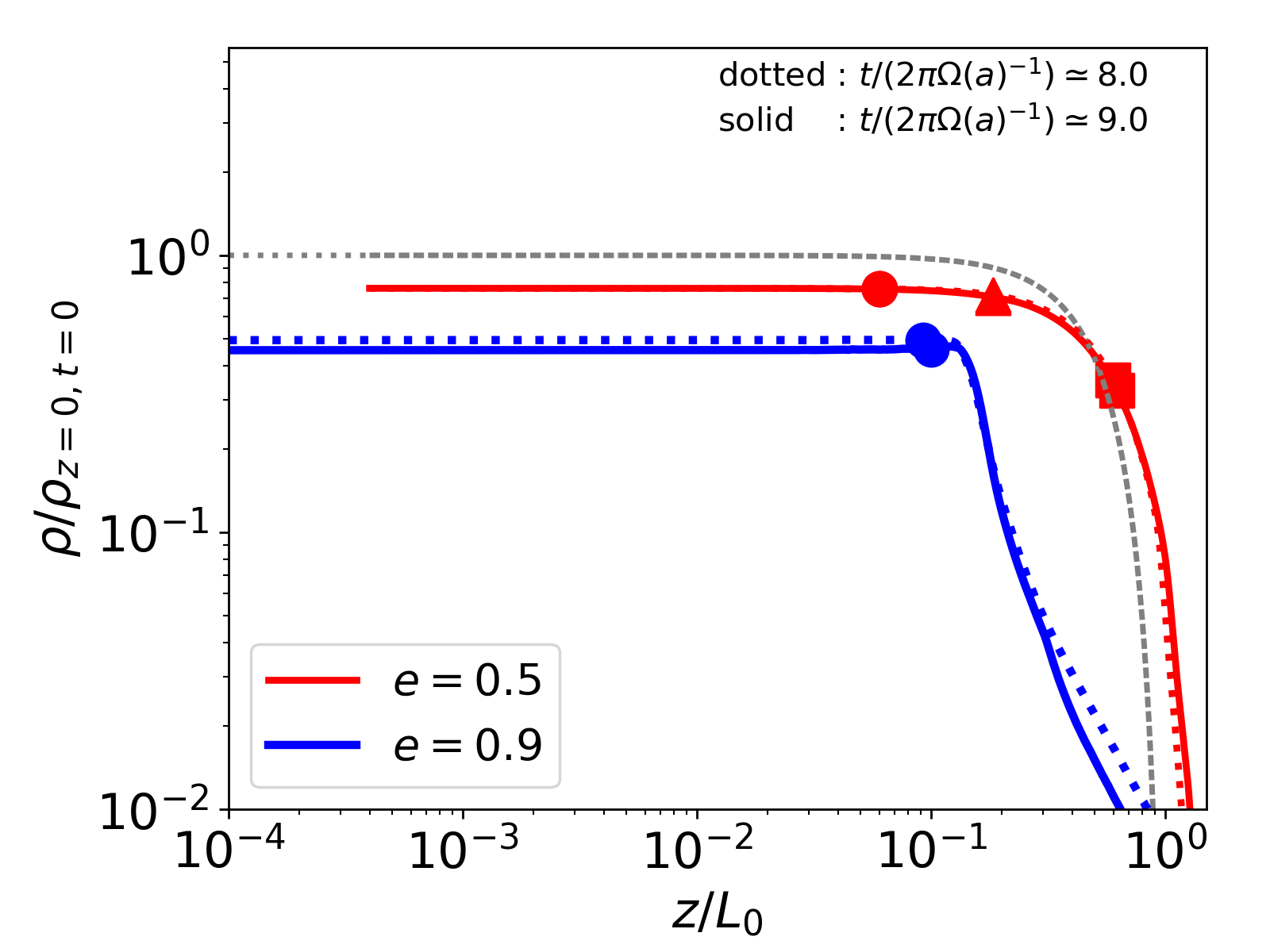}
\caption{Density profiles for $e=0.5$ (red) and $0.9$ (blue) in model \textbf{H01} for two adjacent obits (8th and 9th orbits) when the column is most compressed near pericenter (\textit{top} panel) and expanded near apocenter (\textit{bottom} panel). Here $L_{0}$ is the initial disk thickness. Three values of $M/M_{{\rm tot},t=0}$ are shown with symbols: $M/M_{{\rm tot},t=0}=0.1$ (circle), $0.3$ (triangle) and $0.8$ (square). In both panels, the gray dotted line indicates the initial density profile. }
	\label{fig:scaleheights}
\end{figure}

 Because the temperature is so much higher in the shocked than in the unshocked regions, the column structure is cleanly divided into two distinct segments, one retaining its original density profile, the other much more extended.  A structure with this qualitative character is achieved immediately after the first shock passes through. Figure~\ref{fig:scaleheights} displays the density profiles of two cases (\textbf{H01} $e=0.5$ and $e=0.9$) near pericenter (\textit{top} panel) and apocenter (\textit{bottom} panel) during the limit-cycle. For $e=0.5$, the density profile at apocenter resembles the initial state. By contrast, the density profile for $e=0.9$ changes sharply at the boundary between the shocked and unshocked segments ($z/L_{0}\simeq 6\times10^{-4}$ at pericenter and $\simeq 0.2$ at apocenter). At both pericenter and apocenter, the structure below the break is similar to the initial profile, but above the break the density follows a declining power-law.  The shocked gas is so hot the column's height is almost comparable to $L_{0}$ even though the column mass for $e=0.9$ at these orbits is only 23--25\% of the initial total mass (cf., 98\% for $e=0.5$).For all eccentricities, the column is strongly compressed at pericenter: by roughly one order of magnitude for $e=0.5$, by $\simeq 2.5$ orders of magnitude for $e=0.9$. Lastly, we note that the nearly identical profiles at the two adjacent orbits shown for both high and low-$e$ disks demonstrate that this state is a genuine periodic limit-cycle.

A corollary of the strong shock heating of the outer gas is that the initially uniform entropy distribution becomes increasingly bimodal as the outer part of the disk is repeatedly elevated to a higher entropy state by shocks. For strong shock cases, the entropy in the shocked regions (as measured by the proxy $P/\rho^\gamma$) increases by 1-3 orders of magnitude  compared to the initial value.  In fact, the shocked mass fraction numbers displayed in Figure~\ref{fig:shocked_gas_mass} are, in effect, a representation of the importance of this entropic contrast: the fraction of mass in the column that has been shocked is the fraction of mass with elevated entropy.

\begin{figure}
	\includegraphics[width=8.6cm]{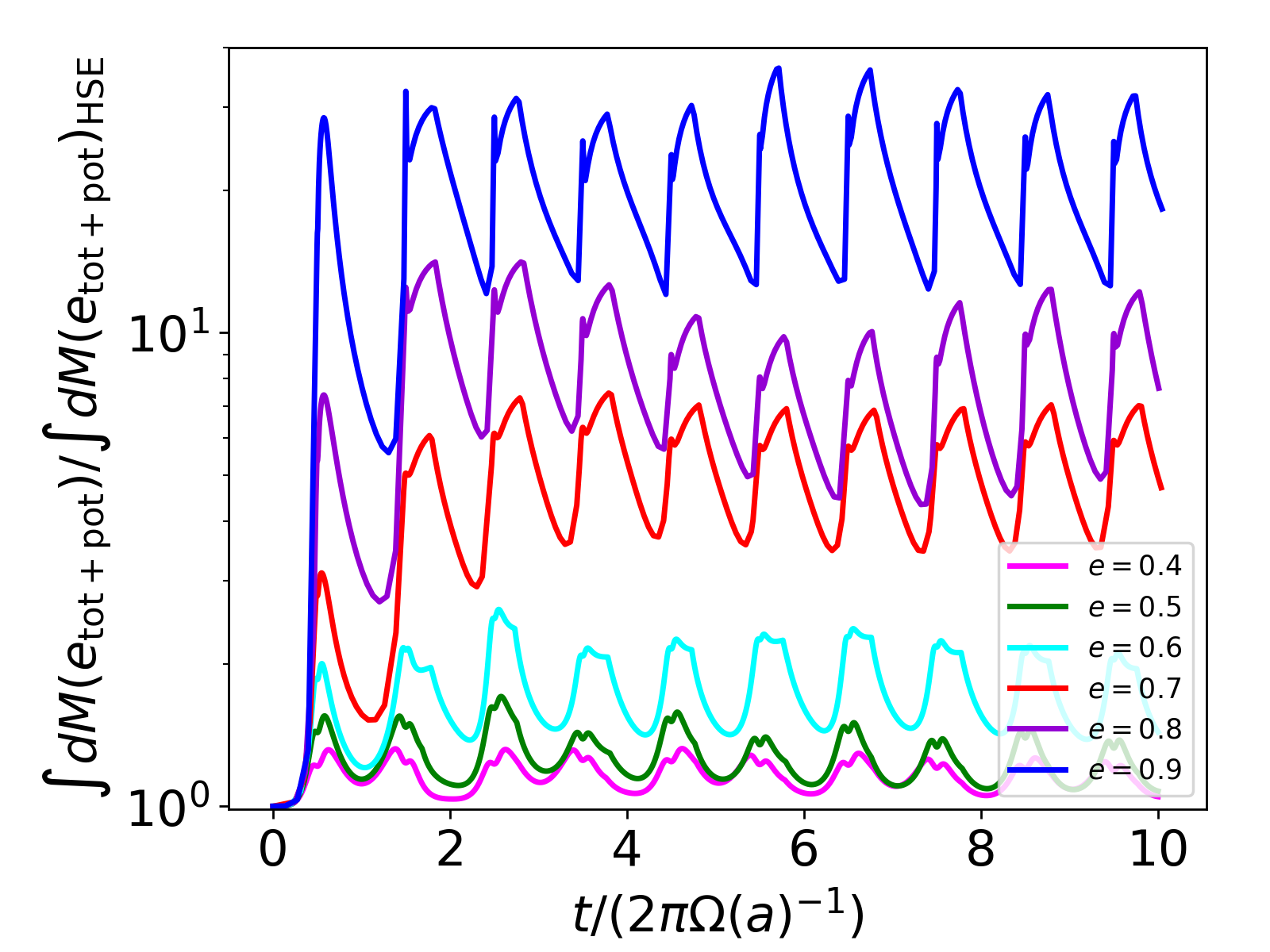}
	\caption{The ratio of actual column-integrated energy to the energy of the corresponding adiabatic (i.e., $K=K_{t=0}$) and hydrostatic equilibrium in the model \textbf{H01}.}
	\label{fig:energy}
\end{figure}

\subsection{Energetics of the shocked gas}\label{subsec:energetics}

The total energy ($e_{\rm tot}+e_{\rm pot}$) evolution of the column is illustrated in Figure~\ref{fig:energy}.
In our model,  posed in the orbital frame,
energy in a column is not conserved because the potential is time-dependent.  From a different perspective, that of a true global disk, non-conservation of individual column energy can be seen as due to our assumption of rotation on (elliptical) cylinders; in a physical disk, this would be enforced by pressure forces within the disk; these forces, which are inserted ``by hand" in our model, can exchange energy between columns.

As a column travels from apocenter to pericenter, it gains energy relative to the instantaneous hydrostatic equilibrium energy, and then loses most of its gain on the trip out to apocenter.  The curves for $e \leq 0.6$ in Figure~\ref{fig:energy} illustrate this behavior.  However, as the eccentricity increases beyond this level, the additional energy near pericenter also becomes greater, and the energy remaining upon return to apocenter becomes much larger: 
Within 2--3 orbits, the column energy per unit mass at apocenter has risen by a factor that increases from $\simeq 1.5$ for $e=0.4$ to $\simeq 15$ for $e=0.9$.  At later times, the specific energy for all eccentricities varies roughly periodically, over a range of about a factor of 2.  Thus, despite the considerable changes in structure between our hydrostatic initial condition and the long-term state, all cases ultimately fall onto a stable limit cycle.
The height of the peak in energy for high eccentricity is due to the deep gravitational potential encountered at the pericenter, a factor $(1+e)/(1-e)$ deeper than the potential at apocenter.  

If there were no dissipation, the gas thermodynamics would be reversible: thermal energy associated with adiabatic compression en route to pericenter would be lost in adiabatic work en route to apocenter. Long-term evolution therefore demands thermodynamically irreversible processes.
Shocks are one such mechanism: they place gas on a higher-entropy adiabat. When shocked gas returns to apocenter, it retains a larger thermal energy, and this greater thermal energy causes it to expand higher than in the initial state.  As will be discussed in Section~\ref{sec:expelled}, there is also a mechanism causing irreversible heat loss: expulsion of matter from the column.

An indicator of the possible importance of mass-loss arises from
another way in which to calibrate the change in energy of the column: to compare $e_{\rm tot}$ to the specific binding energy for matter in the orbital plane, $e_{\rm b} = GM/(2a)$.  For $e=0.9$ of the model \textbf{H01}, the increase in column energy is $\simeq 0.2 e_{\rm b}$ after one orbit and reaches $\simeq e_{\rm b}$ on the limit cycle. 
 This amount of energy is small compared to the potential energy available at pericenter (for $e=0.9$, $20 e_{\rm b}$), but comparable to the potential energy separating the base of the column (i.e., $z=0$) from infinity when it is at apocenter, $1.05 e_{\rm b}$ for $e=0.9$.

  Because the dissipated energy must ultimately be drawn from the orbital energy, and, as we have just seen, for high-eccentricity orbits the amounts in question are comparable to the initial orbital energy, these orbits must evolve as a result of the shock dissipation, diminishing in semimajor axis and becoming less elliptical.  This conclusion can be avoided only if some other mechanism restores the lost orbital energy.

\begin{figure}
	\includegraphics[width=9cm]{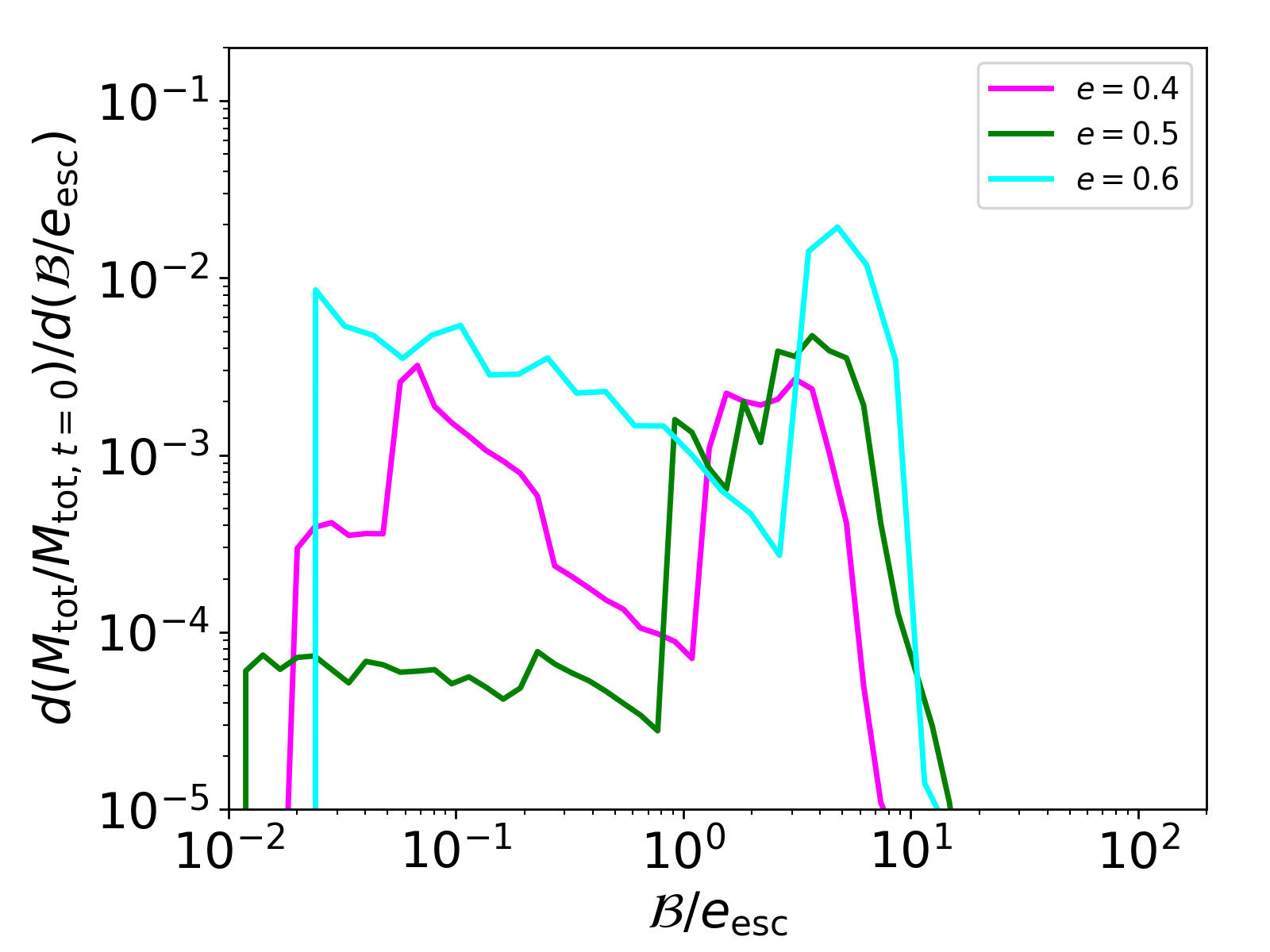}	
	\includegraphics[width=9cm]{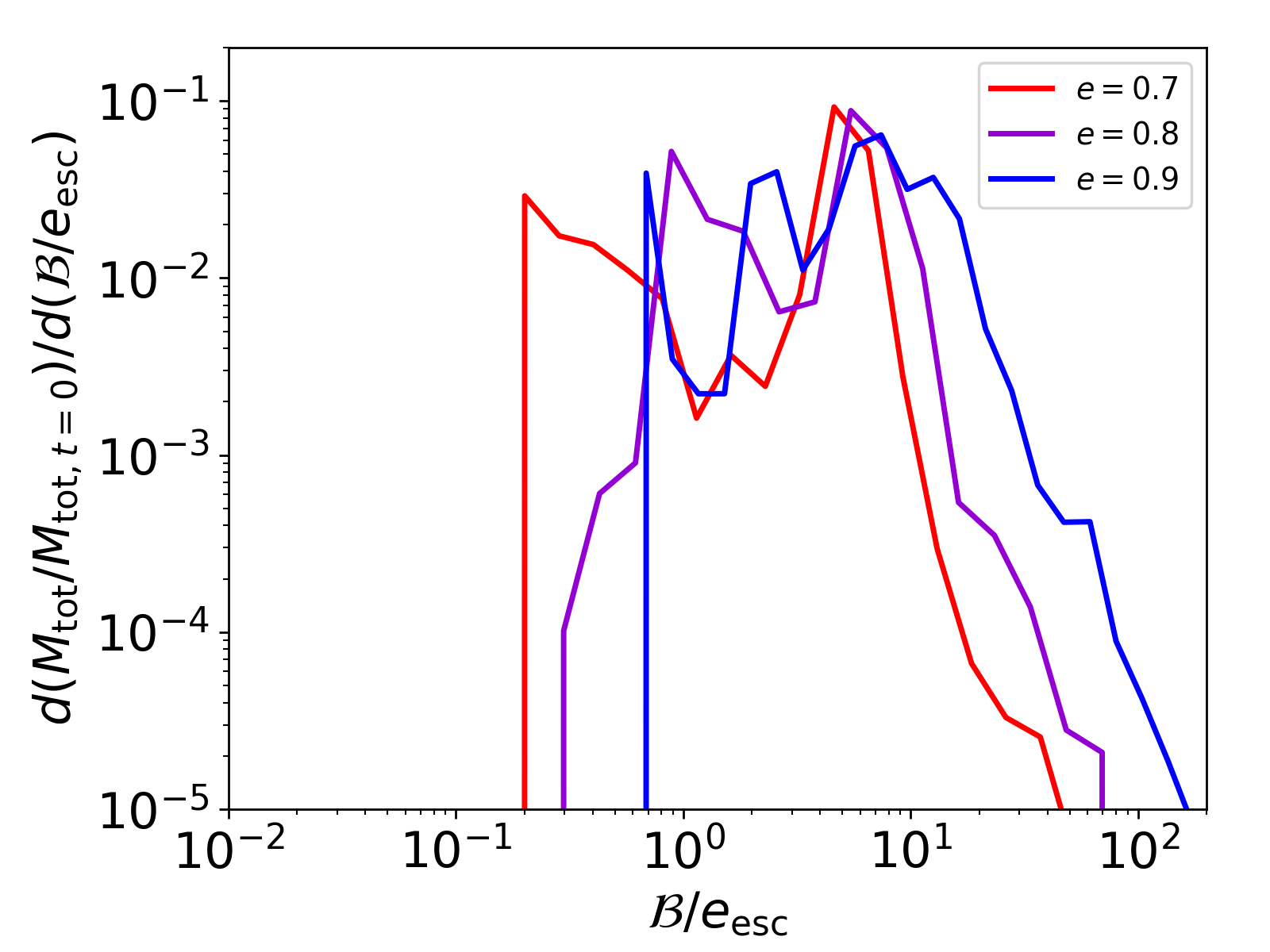}
\caption{The Bernoulli parameter $\mathcal{B}$ distribution of the gas that is expelled from the domain up to ten orbits for $0.4\leq e\leq 0.6$ (\textit{upper} panel) and $0.7\leq e\leq 0.9$ (\textit{bottom} panel). }
	\label{fig:Bernoulli}
\end{figure}

\subsection{Expelled gas}\label{sec:expelled}

A significant fraction of the shock-heated gas can be given so much additional energy that it is expelled supersonically from the domain when the column returns to apocenter ($\mathcal{M} \sim 1-2$ for most of the expelled mass).
However, as shown in the \textit{bottom} panel Figure~\ref{fig:shocked_gas_mass}, the fraction of the column mass expelled per orbit is a very strong function of eccentricity.   At low eccentricity, hardly any is expelled until shocks have begun to heat a small amount of gas at the top of the column.  By extending this gas's scale height, later shocks are strengthened, and after 2--3 orbits, the system reaches the limit-cycle, in which a non-zero, but rather small ($\sim 10^{-3}$ for $e=0.4 - 0.5$), fraction is expelled per orbit.  As can be seen by comparing the two panels of Figure~\ref{fig:shocked_gas_mass}, the expelled gas for these cases is a constant fraction of the gas shocked each orbit, 
 10\% for $e=0.6$ and diminishing to $\approx 3\%$ for lower eccentricities. Higher eccentricity cases ($e \gtrsim 0.7$), reach their long-term expelled fraction almost immediately, and it is much larger: $\gtrsim 0.1$ for $e=0.9$.  This is $\approx 20\%$ of the gas shocked in the orbit. Because the expelled fraction is almost constant, the mass in the column decreases roughly exponentially with time on the limit cycle, but much more rapidly for $e \gtrsim 0.7$ than for less eccentric orbits.

To obtain some idea of what happens to the expelled gas after it leaves the simulation domain, we compute the gas's Bernoulli constant $\mathcal{B} = (1/2)(v_{\rm orb}^2 + v_z^2) + \gamma/(\gamma-1) P/\rho - GM/\sqrt{r^2+z^2}$ at the instant when the mass passes through the outer boundary of the simulation.  
Here, $v_{\rm orb}$ is the velocity of the in-plane motion. When $\mathcal{B} > 0$, the fluid is unbound from the system.  The distribution of ejected mass with respect to $\mathcal{B}/e_{\rm esc}$, where  $e_{\rm esc} = GM/\sqrt{r^2+z^2}$, is shown in Figure~\ref{fig:Bernoulli}.
For all eccentricities, nearly all the expelled mass has $\mathcal{B} > 0$.  For $e\lesssim 0.6$ (the \textit{upper} panel), most of it is unbound with $\mathcal{B}/e_{\rm esc} \lesssim1$, but for $e > 0.6$ (the \textit{lower} panel), most of the expelled gas has a much larger Bernoulli constant: $\simeq 1-10$.  For higher $e$, the distribution also has a longer tail toward larger $\mathcal{B}/e_{\rm esc}$, with 1\% of the mass reaching values of $\simeq10$ for $e=0.7$ and $\gtrsim 23$ for $e\geq0.8$. 

Expelled mass with $\mathcal{B}<0$ is found for $e\leq0.6$, but it is very small. The bound mass fraction relative to the total expelled mass is $3\times 10^{-5}$ for $e=0.4$, $2\times10^{-3}$ for $e=0.5$ and $0.02$ for $e=0.6$; the Bernoulli constant for bound expelled mass is a fraction of $e_{\rm esc}$: $-\mathcal{B}\simeq0.05-0.3$. 

Thus, the gravitational pumping underlying the shocks puts enough energy into a fraction of the column's gas to unbind it. As noted above, long-term stability in the specific energy of the column demands that the heat injected by shocks be balanced by the heat ejected in expelled mass; without radiation, there is no other way to remove heat from the column.  This fact permits an estimate of the shock heating rate per orbit: $\Delta E \approx {\cal B} \Delta M_{\rm expelled}$.
Identifying ${\cal B}$ with the injected heat is justified on the grounds that ${\cal B}$ is conserved along streamlines in adiabatic flow and its value post-shock is in general much larger than it had been pre-shock.  Because ${\cal B}$ is a weakly-growing function of $e$, the shock-heating rate is a slightly stronger function of eccentricity than $\Delta M_{\rm expelled}/M_{\rm tot}$.  Particularly for $e \gtrsim 0.7$, this rate of energy dissipation is quite substantial: the mass-weighted $\langle {\cal B} \rangle \sim 10 e_{\rm esc} \sim 10 e_{\rm b}$ (see Figure~\ref{fig:Bernoulli}), so that $\Delta E \sim 10 e_{\rm b} M_{\rm expelled}$ for these higher eccentricities.

Although the expelled gas is formally unbound, and nearly all of it by a large amount, our simple 1D model is inadequate to support a strong statement about where it goes; in principle, it is possible for it to travel quite far from the disk, but interaction with gas ejected from neighboring columns could strongly affect the expelled gas's fate.

\begin{figure}
\centering
		\includegraphics[width=8.6cm]{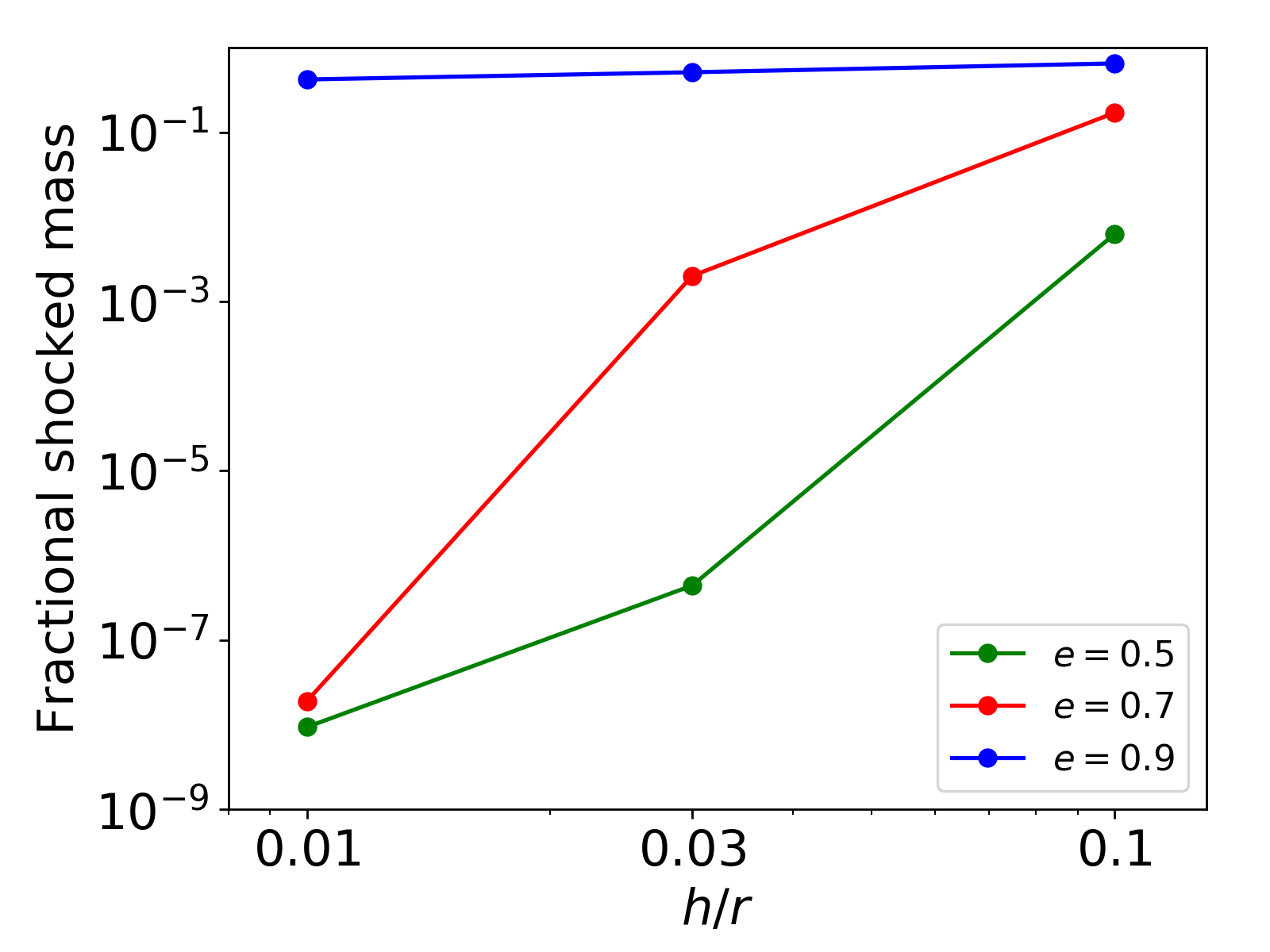}	
		\caption{The fractional mass of shocked gas $M_{\rm{shock}}/M_{{\rm tot},t=0}$ as a function of $h/r$ for $e=0.5$ (green), $0.7$ (red) and $0.9$ (blue). This quantity is evaluated at the end of the first obit.}
	\label{fig:e_L_rapo}
\end{figure}

\subsection{Dependence on the initial thickness}\label{subsec:initialwidth}

Figure~\ref{fig:e_L_rapo} portrays the joint dependence of the shocks' strength on $e$ and $h/r$. 
Compared to the thickest disk models, the shocks created in the thinner models are relatively weak when $e \leq 0.7$. Almost no gas is shocked in the two thinner disk models for $e=0.5$ and $0.7$, but a substantial mass-fraction is shocked for all $e > 0.6$ for the thickest model, \textbf{H01}.  On the other hand, for $e=0.9$, the shocked mass-fraction is significant even in \textbf{H001}---$\simeq 42\%$ of the initial total mass---and increases further with greater intrinsic disk thickness.
Thus, the importance of these shocks is quite sensitive to $h/r$ for eccentricities $\lesssim 0.8$, but for $e \geq 0.9$, the dependence on $h/r$ is weaker.

It therefore appears that, for $e \gtrsim 0.8$, there is no particular barrier preventing even those disks created in a very thin configuration from rapidly heating up to the limit-cycle state found in the \textbf{H01} simulations. On the other hand, the very small gas mass fractions shocked during the first orbit for smaller values of $e$ and initial $h/r \leq 0.03$ suggest that even if these cases have a long-term evolution, they will remain nearly-adiabatic (i.e., following limit-cycles similar to those of the low-eccentricity \textbf{H01} cases) for a great many orbits.

\begin{figure}
\includegraphics[width=8.6cm]{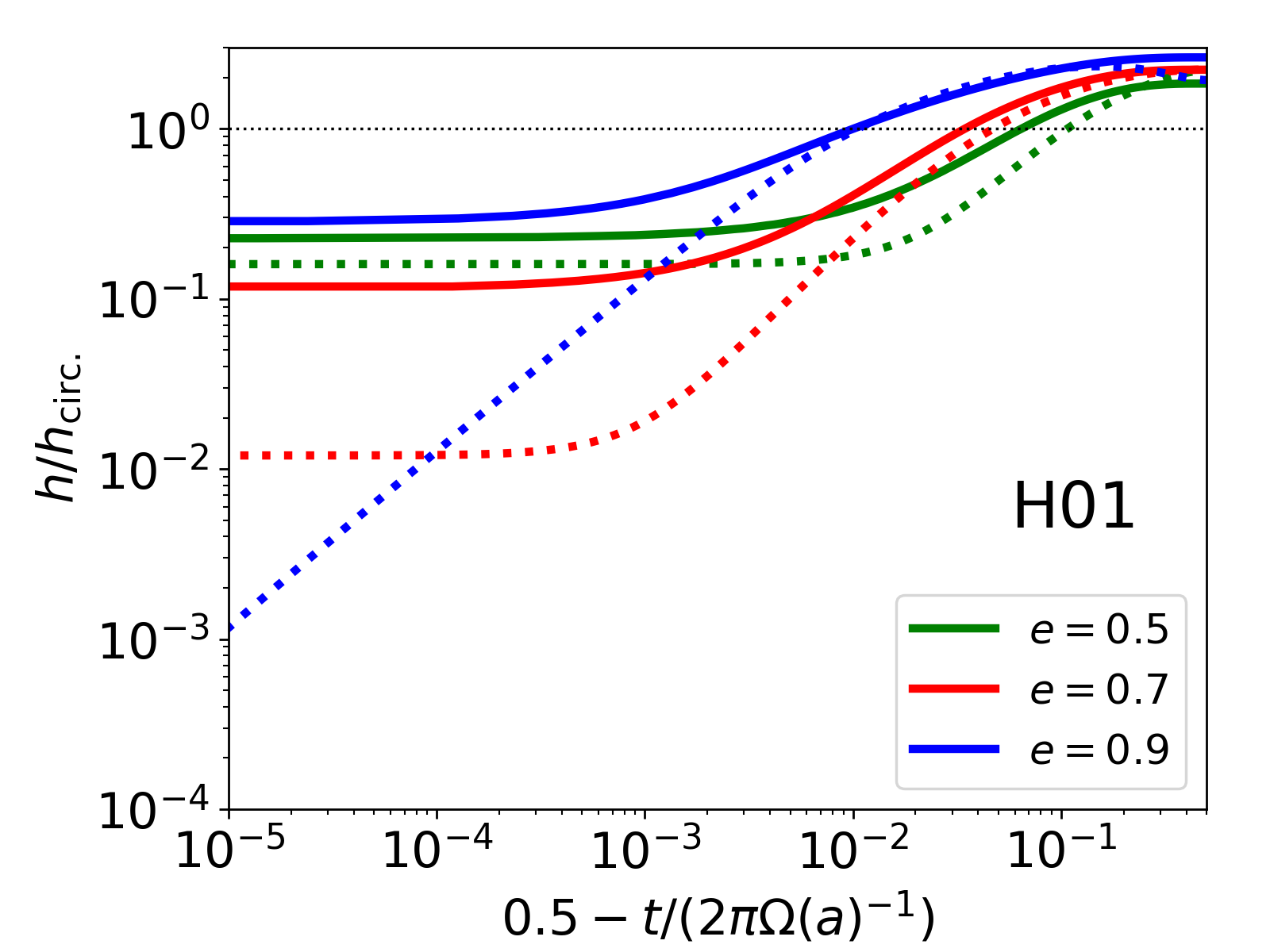}\hspace{-0.15in}
\includegraphics[width=8.6cm]{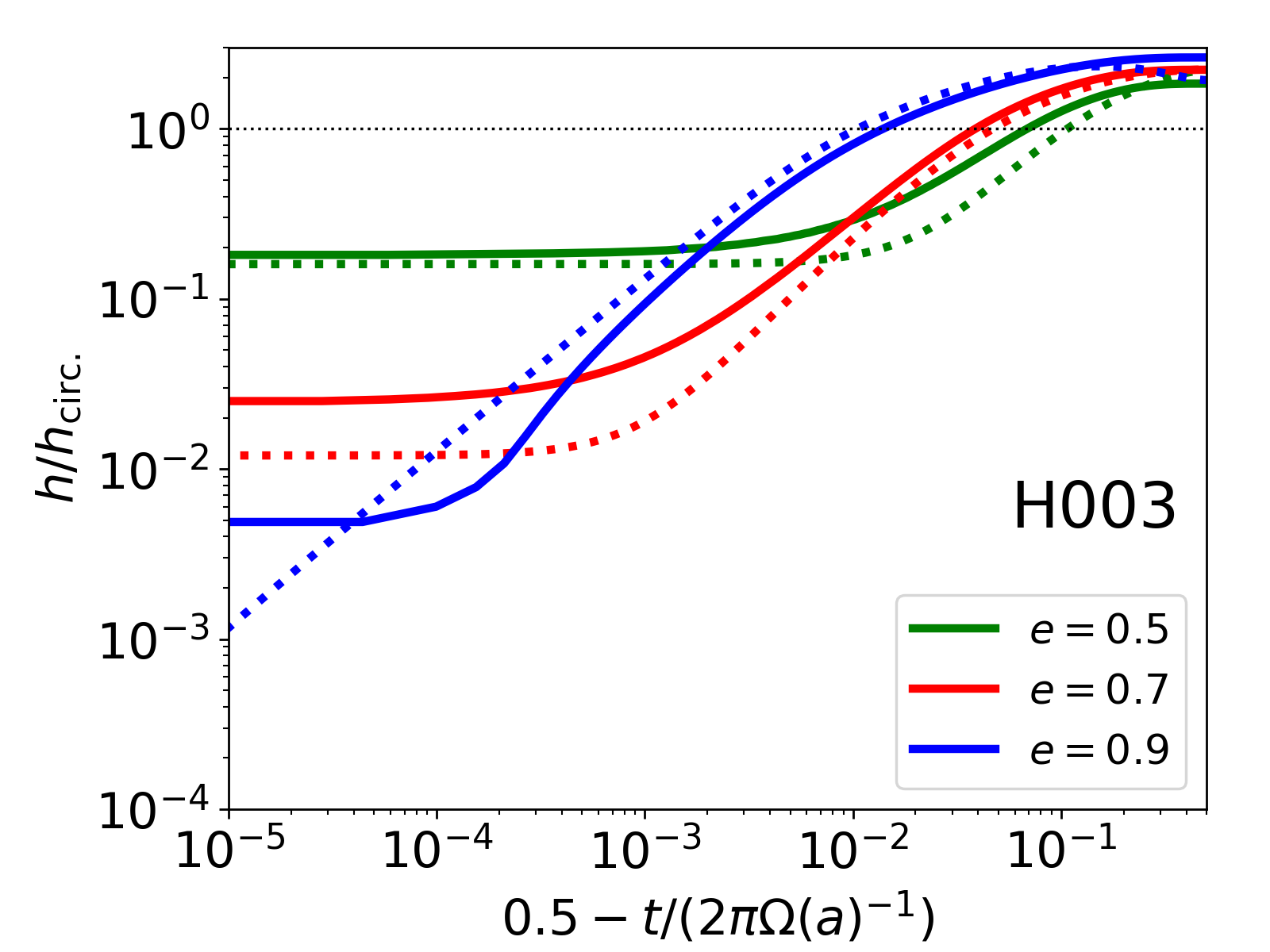}\hspace{-0.15in}
	\caption{The ratio of the density-moment scale height $h$ to that for a circular disk $h_{\rm circ}$. Solid lines represent our models for $e=0.5$, $0.7$ and $0.9$; dotted lines show the analytic solutions of \citet{ZanazziOgilvie2020} for the same values of $e$. We solve Equation 35 in \citet{ZanazziOgilvie2020} using a fourth-order Runge-Kutta method.}
	\label{fig:ZanazziOgilvie}
\end{figure}

\section{ Discussion}\label{sec:discussion}
\subsection{Dynamical effects}\label{subsec:dynamicaleffect}
We have found that fluid columns following eccentric orbits can be strongly altered by  shocks in as little as a single orbit. Particularly for the highest eccentricities and moderately large initial aspect ratios, large fractions of their initial mass can gain a great deal of heat, and some of this heated gas can be expelled.

After only a small number of orbits, they can, as a result of these effects, find their way onto a limit-cycle.  Moreover, the strong contrast in column structure between the initial state we chose (homogeneous in entropy, hydrostatic) and the limit-cycle structure (large internal entropy contrasts, internal shocks, and constant compression or expansion) suggests that the limit-cycle's radius of convergence in phase space may be large.  In fact, for the highest eccentricities, shock heating is strong enough for initially very thin disks to reach this limit-cycle.

During a single period of the limit-cycle (one orbital period), the energy per unit mass of the column varies over a range of a factor 2--3 when $e \gtrsim 0.7$, and its cycle-averaged specific energy can be an order of magnitude greater than its initial value.   Because the column is reheated by a shock every time it passes through pericenter, the energy-balance of the column is maintained by mass-loss if radiative losses are minor.  The fraction of the column mass lost per orbit is $\sim 0.1$ for $e \gtrsim 0.7$, so the longevity of such a system is limited.

The lifetime of the limit-cycle may also be limited in another way.  The heat deposited in the column by the repeated shocks is drawn from the orbital energy.  For the orbit to be maintained therefore requires energy input from some other source.  Without energy resupply, the semi-major axis and eccentricity must decay over time, although eccentricity change is comparatively slow when $e$ is close to unity.  Thus, these results may be interpreted in either of two ways: that highly-eccentric disks, especially those born with moderately large aspect ratios, are intrinsically transient with lifetimes of only a few orbits; or that highly-eccentric disks able to draw upon an orbital energy-replenishment mechanism can follow a limit-cycle for tens of orbits until their mass is depleted by repeated mass expulsions.

Previous work \citep{ZanazziOgilvie2020,LynchOgilvie2021} has also found limit-cycle behavior. For very low eccentricity, particularly in intrinsically thin disks, our results agree with theirs. In Figure~\ref{fig:ZanazziOgilvie}, we compare the ratio of the scale height $h$ to that for a circular orbit $h_{\rm circ}$ in our disk models (\textbf{H01} and \textbf{H003}) for $e=0.5$, $0.7$ and $0.9$ before shocks are created with the corresponding analytic solutions given by \citet{ZanazziOgilvie2020}. The two solutions for weak shocks (i.e., $e=0.5$ in \textbf{H01} and \textbf{H003} and $e=0.7$ in \textbf{H003}) are in good agreement, differing by less than a factor of two.

However, to the extent that shocks matter (high eccentricity, intrinsically thicker disks), they have a very different character. 
In both \citet{ZanazziOgilvie2020} and \citet{LynchOgilvie2021}, the vertical profile varies homologously at all times.  In the model of \cite{ZanazziOgilvie2020}, the column retains its initial state of homogeneous entropy, while in the model of \cite{LynchOgilvie2021}, a phenomenological heating rate balanced by equilibrium radiative diffusion permits evolution to a new entropy state. In strong contrast to their results, as shown in Figure~\ref{fig:ZanazziOgilvie},  we find the column compression $h/h_{\rm circ}$ for $e\geq0.7$ in \textbf{H01} and $e=0.9$ in \textbf{H003} near pericenter is 1 -- 2 orders of magnitude smaller than predicted by \citet{ZanazziOgilvie2020} at a time shortly before the shocks form (see also Fig.~\ref{fig:vertical_velocity}).

The fundamental reason for the contrast between our results and theirs is that, by not assuming homologous behavior, our hydrodynamic calculations  provide full information about the effects of pressure gradients.  This diminishes compression even in the adiabatic regime, and makes it possible to find solutions with shocks. These effects cannot arise in the previous models because their assumption that all physical quantities' variations can be factored into a product of one function of $z/h$ and another function of $t$ excludes any behavior involving finite sound speeds: such behavior rests fundamentally on functional forms involving $z - (v\pm c_s) t$, which cannot be contained in such a factorization. Even without the factorization assumption, these models also prevent shocks by requiring homogeneous entropy, even if its level changes over time.

 Three contrasts in the nature of the limit-cycle found are especially noteworthy. All three are attributable to the greater sensitivity to pressure forces in our hydrodynamic method.  One is the compression we find as the column travels from apocenter to pericenter is large, but is nonetheless much smaller than found in this previous work.  For example, the apocenter/pericenter compression factor (central density ratio) we find on the limit-cycle for $e=0.9$ is $\approx 400$ (Figure~\ref{fig:scaleheights}), whereas it is $\approx 2 \times 10^5$ in the periodic solution of \citet{ZanazziOgilvie2020}.
The second is our very different entropy distribution. The gas column has a low-entropy region near the midplane, but a much higher-entropy region extending to such heights that, near apocenter, significant gas can be lost. In both of the earlier solutions, the entropy is homogeneous and mass-loss is not considered.  The third is the mass ejection, in which a constant fraction of the shocked gas is lost every apocenter passage.   Although this fraction is small for low-eccentricity cases, it is always present at some level, and is very significant for high-eccentricity.

\subsection{Radiation effects}\label{subsec:genericeffect}

Some of the dynamical conclusions just discussed may be altered when additional physical processes are considered.  For example, in the service of simplicity, our simulations ignore radiation.  Real disks, of course, do not.  Specific predictions about the effects of radiation demand detailed calculations.  Here we limit ourselves to a qualitative discussion of the possible varieties of behavior caused by radiation effects.

In optically thick disks, diffusing photons { smooth the disk's temperature profile on a timescale $t_{\rm cool} =  (1+e_{\rm gas}/e_{\rm rad})t_{\rm esc}$}; $t_{\rm esc}$ is the time for photons to  cross a length $T/|\nabla T|$ by random walk. Here, $e_{\rm gas}$ and $e_{\rm rad}$ are the gas and radiation contributions to the internal energy density.
The cooling time may also be defined locally:
\begin{align}
t_{\rm cool} = \frac{\rho(e_{\rm rad} + e_{\rm gas})}{dF_{\rm diff}/dz},
\end{align}
where  the diffusive flux $F_{\rm diff}=[4~a_{\rm rad}~c T^{3}/(3\kappa\rho)](dT/dz)$ if radiation-gas energy-exchange is rapid enough to thermalize the radiation (in this context $T$ is the true thermodynamic temperature).
However, for full-column considerations only the shock-heated region is relevant because its heat content completely dominates the total.

In addition to regulating the observed luminosity, photon diffusion can also affect column hydrodynamics by altering the pressure.  The degree to which it does so depends on the ratio of $t_{\rm cool}$ to the relevant evolutionary timescale. 
Our simulations correspond to the limit in which { $t_{\rm cool}$  is larger than all evolutionary timescales} at all times.

The qualification ``at all times" is significant to eccentric disks because $t_{\rm cool}$ itself is a function of time.  Even if the opacity is independent of density and temperature so that the fixed surface density of a column makes its optical depth constant, $t_{\rm esc} \propto h$, and the changing vertical gravity induces changes in $h$ that can be quite large.  Thus, depending on system parameters, the long diffusion time limit may apply fully, partially, or not at all.

The simplest case is the long diffusion time limit.
In the truly long diffusion time limit, only a small fraction of the heat escapes, and the disk's temperature profile (with respect to mass) remains unchanged except for mass-loss from the end of one shock to the beginning of the next.  When the next shock passes through, the gas can be heated further.
The only modification to the temperature profile due to radiation is then a downturn at the very top of the hot layer, where the optical depth is small enough that the photon escape time is less than the shock recurrence time, i.e., the orbital period.

If $t_{\rm cool}$ is shorter than the orbital period, but longer than a shock passage time,
the shock creates a new temperature profile within the column, but it evolves significantly only after the shock disappears.  As shown in Figure~\ref{fig:T_S}, the temperature profile left behind by the shock generally peaks at the column surface.  Its evolution begins with quick non-diffusive losses near the photosphere, followed by slightly slower diffusive losses from the layer just under the photosphere.  The temperature now diminishes both upward and downward from a peak near the top of the column, so that photon diffusion carries heat away from the peak both outward and inward.  Because the density declines outward, the outward flux is larger than the inward when comparing two locations with equal temperature.  As a result, the peak temperature both falls and moves deeper into the column.

The definition of $t_{\rm cool}$ in this regime requires some refinement.  The instantaneous cooling time for a heated region of vertical dimension $h$ and optical depth $\tau$ is $\sim (1 + e_{\rm gas}/e_{\rm rad})(\tau h/c)$, but $h$ in general changes significantly as the column travels around an eccentric orbit, and $\tau$  can change if the associated temperature and density changes affect the opacity.  Thus, the criterion that $t_{\rm cool}$ is less than an orbital period and greater than a shock lifetime is more precisely defined by the  requirement that most of the heat deposited by the shock is radiated over a timescale in   this range.

Loss of heat by radiation is, like loss of heat by mass expulsion, a thermodynamically irreversible process.  In terms of long-term changes, it can therefore substitute for mass expulsion.  In particular, when the cooling time is less than half an orbital period, it will not stretch so high when near apocenter, suppressing mass expulsion to some degree.  If a limit-cycle can still be maintained, its cycle-averaged specific energy would be reduced, but the exponential decay of its mass would be considerably slowed.

 Photon diffusion becomes especially interesting when $t_{\rm cool}(r_{\rm peri})$ is less than the lifetime of the shock (as noted in Section~\ref{subsec:vertical}, shock lifetimes are very brief compared to an orbital period).  More precisely, when the shock reaches an optical depth from the surface
\begin{equation}
\tau_{\star} = (c/v_{\rm shock})(1 + e_{\rm gas}/e_{\rm rad})^{-1},
\end{equation}
a phenomenon called ``shock breakout" occurs (see the recent review by \citealt{LN2020}). In such an event, the photons created from the shock's energy dissipation escape from the gas before the shock can move significantly.  As a result, the shock weakens and slows, while radiating with a surface brightness $\sim c v_{\rm shock}^2/[\kappa h^\prime(1 + e_{\rm gas}/e_{\rm rad})]$.  Here $\kappa$ is the opacity and $h^\prime$ is the pressure length scale near the critical optical depth.
When $\tau_{\star}$ is comparable to the total optical depth of the shock-heated region, radiative cooling can severely alter the global character of the shock. In this limit of especially rapid cooling, alterations to the dynamical patterns found in our simulations---strong shock-heating, limit-cycle behavior, etc.---would be especially severe.

 Thus, it is difficult to make general statements both about how radiative effects alter dynamical properties of eccentric disks and about their radiative output. Different  ratios of cooling time to dynamical time lead to quite different outcomes, and that ratio depends strongly on disk parameters (for example, when $t_{\rm esc}$ is greater than the orbital period, the radiation transfer solution is time-dependent and the flux is no longer given by Fick's Law, as assumed in \citealt{LynchOgilvie2021}). In certain parameter regimes, particularly those in which the cooling time is comparatively long, clearer statements can be made, but in other cases, more elaborate calculations are essential: time-dependent radiation transport coupled to the hydrodynamics, and possibly global radiation hydrodynamics simulations.

\subsection{Application to tidal disruption events}\label{subsubsec:TDE}

There is, however, a particularly interesting context for which a few estimates have significant implications.  Tidal disruption events (TDEs) are prime examples of astrophysical events likely to create highly eccentric, moderately thick disks. They occur when stars orbit so close to massive black holes that they are torn apart by tidal gravity. Roughly half the debris is bound and the other half is unbound. Energy released as the bound debris falls back toward the BH  powers the TDE flare;  many such flares have been observed so far \citep{Saxton+2008,Gezari+2009,Holoien+2016,vanVelzen+2021}. 
When the orbital pericenter is small enough, strong relativistic apsidal precession causes shocks in the orbital plane that instantly circularize the very eccentric orbits of the debris, creating a compact circular disk \citep{Rees1988}.
However, events in which this happens are likely to be only a small fraction of all TDEs  \citep{Krolik+2020}, and in most events, apsidal precession is rather weak. Energetics considerations, the so-called ``inverse energy crisis" \citep{Piran+2015}, also show that  rapid circularization typically does not happen because the process itself would produce more energy than is typically observed.

On the other hand, shocks can be created in other ways.  Upon the debris stream's first return to pericenter, a ``nozzle" shock is formed when streams following ballistic orbits tilted with respect to each other by angles $\sim 0.01 (M_{\star}/\Msol)^{1/3}(M_{\rm BH}/10^{6}\Msol)^{-1/3}$ converge \citep{EvansKochanek1989,Kochanek1994}.  In the limit of eccentricity very close to unity, negligible pressure support, and disk aspect ratio $h/r \sim 0.01$, the vertical shocks we have studied would become very similar to these nozzle shocks.

In addition, the numerical work by \citet{Shiokawa+2015} showed that {weak apsidal precession creates shocks near apocenter rather than pericenter (see also \citealt{Dai2015})}. Because the energy dissipation by the outer shocks is not sufficient to circularize the debris, the outcome is a flattened, extended eccentric accretion flow. Since then, other numerical simulations have also shown evidence of the formation of eccentric disks \citep[e.g.,][]{Sadowski+2016,BonnerotLu2020}. Motivated by the simulation results, there have been efforts both physical \citep{Piran+2015,Ryu+2020a} and phenomenological \citep[e.g.,][]{Liu+2017, Cao+2018, Zhou+2020} to build models of eccentric disks in TDEs.

Numerical simulations of TDEs \citep[e.g.,][]{Shiokawa+2015,Sadowski+2016,BonnerotLu2020} have also shown that after the debris has passed through the nozzle and apocenter shocks, it takes the form of an elliptical disk whose eccentricity and aspect ratio are high enough for vertical gravitational pumping to create the sort of shocks studied here {(the criteria are illustrated in Figure~\ref{fig:e_L_rapo})}.
TDE disks are therefore likely to become still hotter and more vertically extended.  They may also become sources of rapidly-moving unbound gas near apocenter (Section~\ref{subsec:energetics}).

The energy conversion by these shocks driven by vertical gravity could be a potentially significant energy dissipation mechanism  supplementary to energy dissipation in the outer shocks driven by orbital motion \citep{Shiokawa+2015}.  The outer shocks can heat the gas up to $\sim 0.1\times$ the
virial temperature at apocenter \citep{Shiokawa+2015}; as shown in Section~\ref{sec:expelled}, the vertical shocks can likewise deposit comparable amounts of energy ($\sim e_b$ per unit mass) in disks as thick and eccentric as created in TDEs, thus augmenting the available heat at the order-unity level.

To gain a sense of the regime in which these disks may operate, it is useful to make a few estimates.
Eccentric disks in TDEs are {generally} optically thick: the vertical electron scattering optical depth to the midplane is
\begin{align}\label{eq:tau}
\tau_{\rm T} =&  \frac{\kappa_{\rm T}M_{\star}}{4\uppi a^{2}\sqrt{1-e^2}},\nonumber\\
\simeq& 500 \left(\frac{M_{\rm BH}}{10^{6}\Msol}\right)^{-4/3}\left(\frac{M_{\star}}{M_{\odot}}\right)^{5/9}\Xi^{2}(1-e^{2})^{-0.5},
\end{align}
where $\kappa_{\rm T} = 0.34\cm^{2}\mathrm{g}^{-1}$ is the Thomson opacity. Here we have determined the characteristic semimajor axis of the debris orbits by $a = 0.5 R_{\star} (M_{\rm BH}/M_{\star})^{2/3}$ and have adopted a phenomenological $M_{\star}-R_{\star}$ relation,  $R_{\star}=0.98(M_{\star}/M_{\odot})^{8/9}R_{\odot}$ \citep{Ryu+2020c}. The function {$\Xi=\Xi(M_{\star}, M_{\rm BH})$ is an order-unity correction factor for the debris' energy width accounting for both realistic internal stellar structure and relativistic effects \citep{Ryu+2020a}}, 
\begin{align} 
\Xi = &\frac{0.62+\exp{[(M_{\star}-0.67)/0.21]}}{1 + 0.55~\exp{[(M_{\star}-0.67)/0.21]}}\nonumber \\
&\times \left[1.27 - 0.3\left(\frac{M_{\rm BH}}{10^{6}\Msol} 
\right)^{0.242}\right].
\end{align}

Using this optical depth, we may estimate the cooling time.  Evaluating it in units of the instantaneous dynamical time (and assuming that $e_{\rm rad} \gg e_{\rm gas}$), we find
\begin{align}\label{eq:difftime}
t_{\rm cool}\Omega(r) &\simeq 1.1 \left(\frac{r}{a}\right)^{-1/2}\left(\frac{h/r}{0.1}\right)  \left(\frac{M_{\rm BH}}{10^{6}\Msol}\right)^{-7/6} \left(\frac{M_{\star}}{M_{\odot}}\right)^{4/9}\nonumber\\ 
&\times\Xi^{5/2}(1-e^2)^{-1/2}.
\end{align}
Because these debris disks are expected to be geometrically thick (e.g., the simulation of \citealt{Shiokawa+2015} found that $h/r \simeq 0.5$), the cooling time is typically comparable to or greater than the orbital dynamical time.  The cooling time for radii $\sim a$ is therefore also of order the orbital period, but smaller in ratio by a factor of $2\pi$. In this sense, TDE disks fall into the category of those capable of significant cooling within an orbital period.  However, this diffusion time estimate refers to the entire column, whereas the shock-heated portion has both a shorter height and a smaller total optical depth, so the diffusion time for the heated region alone should often be shorter than an orbital period.  On the other hand, larger eccentricity slows diffusion and increasing $M_{\rm BH}$ accelerates it because $\tau$ is inversely proportional to the ellipse area, which is $\propto M_{\rm BH}^{4/3}(1-e^2)^{-1/2}$.

To compare the cooling time to the shock duration time, several additional adjustments must be made to Equation~\ref{eq:difftime} in order to convert $\Omega(r)$ to the inverse of the shock lifetime.  First, $(r/a)^{-1/2}$ becomes $(1-e)^{-1/2}$ because the shocks occur only when the column is very near pericenter.  Second, $h/r$ changes from apocenter to pericenter; for $e=0.5$, it decreases by a factor $\simeq 2$, whereas for $e=0.9$, it increases by a factor $\simeq 3$.  Third, the shock duration time is generally rather shorter than $\Omega^{-1}(r_{\rm peri})$ for $e \lesssim 0.7$ and is $\gtrsim \Omega^{-1}(r_{\rm peri})$ for $e \gtrsim 0.8$ (see discussion in Section~\ref{subsec:vertical}).  Combined, these three factors tend to make diffusion more rapid when $e$ is small, but slower when it is large. 

 In this context, it is especially important to emphasize that
 $t_{\rm cool}$ for the shocked region is in general shorter than the estimate for the entire column given in Equation~\ref{eq:difftime}.  In addition, as the shock rises upward, the cooling time of immediately post-shock matter diminishes further. It is likely, then, that radiative losses and shock breakout are very significant for lower eccentricity TDE debris disks (these are also the cases in which the thickness of the shocked debris is least), and possibly 
 even for the highest eccentricities ($e \gtrsim 0.9)$.  Cooling processes are also relatively more effective for events involving higher mass black holes.

Thermalization may be marginal because the optical depth required for thermalization,
\begin{align}
\tau_{\rm therm} \simeq \sqrt{\frac{\kappa_{\rm T}}{\kappa_{\rm ff}}}&
\simeq 3\times10^{2}  \left(\frac{M_{\rm BH}}{10^{6}M_{\odot}}\right)^{0.23} \left(\frac{M_{\star}}{M_{\odot}}\right)^{-0.13}\left(\frac{h/r}{0.1}\right)^{0.063} \nonumber\\
&\times\Xi^{0.31} (1-e^{2})^{0.03}(1+e)^{-0.44},
\end{align}
is similar to the total optical depth, although it may be more complete for events with larger $M_{\rm BH}$ because $\tau_{\rm therm}/\tau_T \propto M_{\rm BH}^{1.56}\Xi^{-1.69}$.  Here, $\kappa_{\rm ff}$ is the free-free opacity. Note that we assume the bulk of the disk's surface density has been heated enough to make its ``pressure-support" temperature $\sim 0.1 T_{\rm virial}$.  This assumption determines the fiducial thermodynamic temperature
\begin{align}
    T\simeq & 1.3\times10^{5}  \left(\frac{M_{\rm BH}}{10^{6}M_{\odot}}\right)^{-1/4}\left(\frac{M_{\star}}{M_{\odot}}\right)^{1/12}\Xi^{3/4} \times \nonumber \\
    &\left(\frac{h/r}{0.1}\right)^{-1/4}(1+e)^{-1/4}(1-e^{2})^{-1/8}\hbox{~K}.
\end{align}
Because of the large total optical depth, the effective temperature at the photosphere should be a factor of several smaller.

As these estimates demonstrate, the detailed character of the radiation emitted by the shocked gas in the TDE context is sensitive to parameters.  
The shocks may be seriously weakened by radiation losses in some cases, but not in others. The hot column may or may not cool during intervals between successive near-pericenter shocks, and its emergent spectrum may or may not be close to a black body, although thermalization should be more complete at frequencies below the Planck spectrum's peak.  For these reasons, and because global effects are likely to alter these estimates in meaningful ways, we postpone a more specific treatment of the predicted radiation to later work.

\section{Summary and Conclusion}\label{sec:summary}

As we have shown analytically, eccentric disks cannot maintain hydrostatic equilibrium, and generically drive supersonic vertical motions which lead to shocks. To understand the effects of these motions quantitatively, we have performed
$1D$ hydrodynamics simulations for isolated columns of gas in
eccentric disks. These simulations revealed that the immediate importance of shocks to disk evolution depends strongly on the intrinsic disk thickness and the eccentricity of the disk's orbits. When $h/r \gtrsim 0.1$, a sizable fraction of a disk column's mass is shocked within a single orbit if $e \gtrsim 0.6$; when $h/r \lesssim 0.03$, the threshold eccentricity for such strong shocks is $e \simeq 0.8$. In disks intrinsically thin enough and having small enough eccentricity to make the shocks weak, a simpler model in which columns evolve homologously \citep[e.g.,][]{ZanazziOgilvie2020,LynchOgilvie2021} is a good approximation.

The energy for the supersonic motions, and therefore for the shocks, is injected by gravitational forces near pericenter, where the gravitational potential is a factor $(1+e)/(1-e)$ deeper than at apocenter.  Without the shocks, these gravitational forces acting on disk matter would do positive work in some locations of the orbit and negative work in others, so that the net deposited energy is small. 
However,
the heat injected into the gas by shocks raises its entropy, irreversibly placing the gas on an elevated adiabat.
The ``loan" of energy to matter when it travels through the deeper part of the potential is thus converted into a permanent ``gift" by shock dissipation. Depending on event parameters, this gift may be either kept or discarded, the latter option occurring when photon diffusion is sufficiently rapid.

When $e$ is comparatively large, the energy per unit mass available near pericenter is much larger than the escape energy near apocenter.  This contrast makes it possible for the dissipated energy to be a significant fraction of the orbital binding energy. Some of the heated gas can then become unbound and expelled when carried out to apocenter.  

After a few orbits, the columns begin to traverse a limit-cycle in which the net specific energy change over a cycle period is nearly zero.
When the shocks are weak, the evolution is nearly adiabatic. The limit-cycle is then a nonlinear ``breathing'' mode in which the time-varying potential pumps in energy through compression, and this energy is returned as the column expands.  On the other hand, when the shocks are strong, the specific energy of the column can be changed drastically from its initial state, and its thermodynamics become controlled by non-adiabatic processes: heat dissipated in shocks is vented by the expulsion of hot gas near apocenter.  In this case, the outer layers of the column have much higher entropy than the midplane region, which retains its initial entropy. That our hydrostatic equilibrium initial condition can be transformed into a much hotter and strongly time-dependent limit-cycle in 2--3 orbits indicates that the limit-cycle is a generic solution.

However, the system is intrinsically transient in two ways. On the limit-cycle, the column loses a constant fraction of its mass every orbit, a large fraction at high eccentricity, a much smaller fraction at low eccentricity.  In addition, the ultimate source of the energy used to heat the disk gas is the orbital energy of the disk.  Thus, dissipation in shocks must lead to diminishing the disk's semimajor axis and, to an extent that diminishes with increasing $e$, its eccentricity.   Although invisible to this 1D simulation in which the matter is forced to follow a given orbit, the decay of both semimajor axis and eccentricity should be apparent in global simulations of highly-eccentric disks unless some additional mechanism restores the energy.

Depending on circumstances, radiation effects may alter column structure and dynamics.  The immediate post-shock state possesses an inverted temperature gradient, but radiative diffusion can erode its peak temperature and restore the usual sign of the gradient.
If the cooling time is shorter than the orbital time, radiation losses can remove the heat deposited by shocks before the next return to pericenter. If the cooling time is shorter than the shock propagation time, the shocks may be greatly weakened.  

A prime example of disks both geometrically thick enough and eccentric enough for vertical shocks to be important are those created by TDEs 
\citep{Shiokawa+2015, Sadowski+2016, BonnerotLu2020}.  In these disks,
the radiative diffusion time at apocenter is comparable to the orbital period, which is the interval between successive shocks, and the diffusion time near pericenter may be comparable to or shorter than the duration of the shocks. These two facts taken together make it especially important for further studies to compute time-dependent radiation transfer in coordination with the hydrodynamics.

To conclude, we have found a
 channel for shock-driven energy transfer from orbital energy to internal energy that can lead to dramatic changes in the vertical structure and orbital properties of eccentric disks. Strong shocks occur in thicker and more eccentric disks. Such conditions may be found in interesting astrophysical contexts such as the accretion flows resulting from tidal disruption events.  {In those cases, it may provide a significant supplement to the dissipation due to previously-proposed mechanisms.}
Because the most dramatic impacts of these shocks also lead to breakdowns in this simple picture of independent gas columns, new phenomena can be expected when they are placed in the context of global $3D$ dynamics. Examples include: changes in the orbital plane velocity due to horizontal pressure gradients, interactions between streams of gas expelled from different columns, orbital evolution, obliquity in the shocks whose $1D$ character we have explored here.

\section*{Acknowledgements}
The authors thank the anonymous referee for the constructive comments and suggestions. Also, we thank Elliot Lynch for his comments which help us improve the manuscript.  JHK and TR were partially supported by NSF grant AST-1715032. TP was supported by an advanced ERC grant TReX. 
 The authors acknowledge the analysis toolkit \texttt{yt} \citep{Turk+2011} and matplotlib \citep{Hunter:2007} for making the plots in the paper. This project was conducted using computational resources (and/or scientific computing services) at the Maryland Advanced Research Computing Center (MARCC). The authors would also like to thank Stony Brook Research Computing and Cyberinfrastructure, and the Institute for Advanced Computational
Science at Stony Brook University for access to the high-performance
SeaWulf computing system, which was made possible by a $\$1.4$M National Science Foundation grant (\#1531492).

\software{
\small{CASTRO} \citep{Almgren+2010}; \texttt{yt} \citep{Turk+2011}; matplotlib \citep{Hunter:2007}.
}

\appendix
\section{Convergence test}\label{appendix:res}
To show the convergence of our simulations, we performed simulations with different resolutions for the fiducial model \textbf{H01}  with $e=0.5$, $0.7$ and $0.9$. For each $e$, we consider lower and higher resolutions by a factor of 1.5 relative to that for the fiducial model. 

\begin{figure*}
\includegraphics[width=8.5cm]{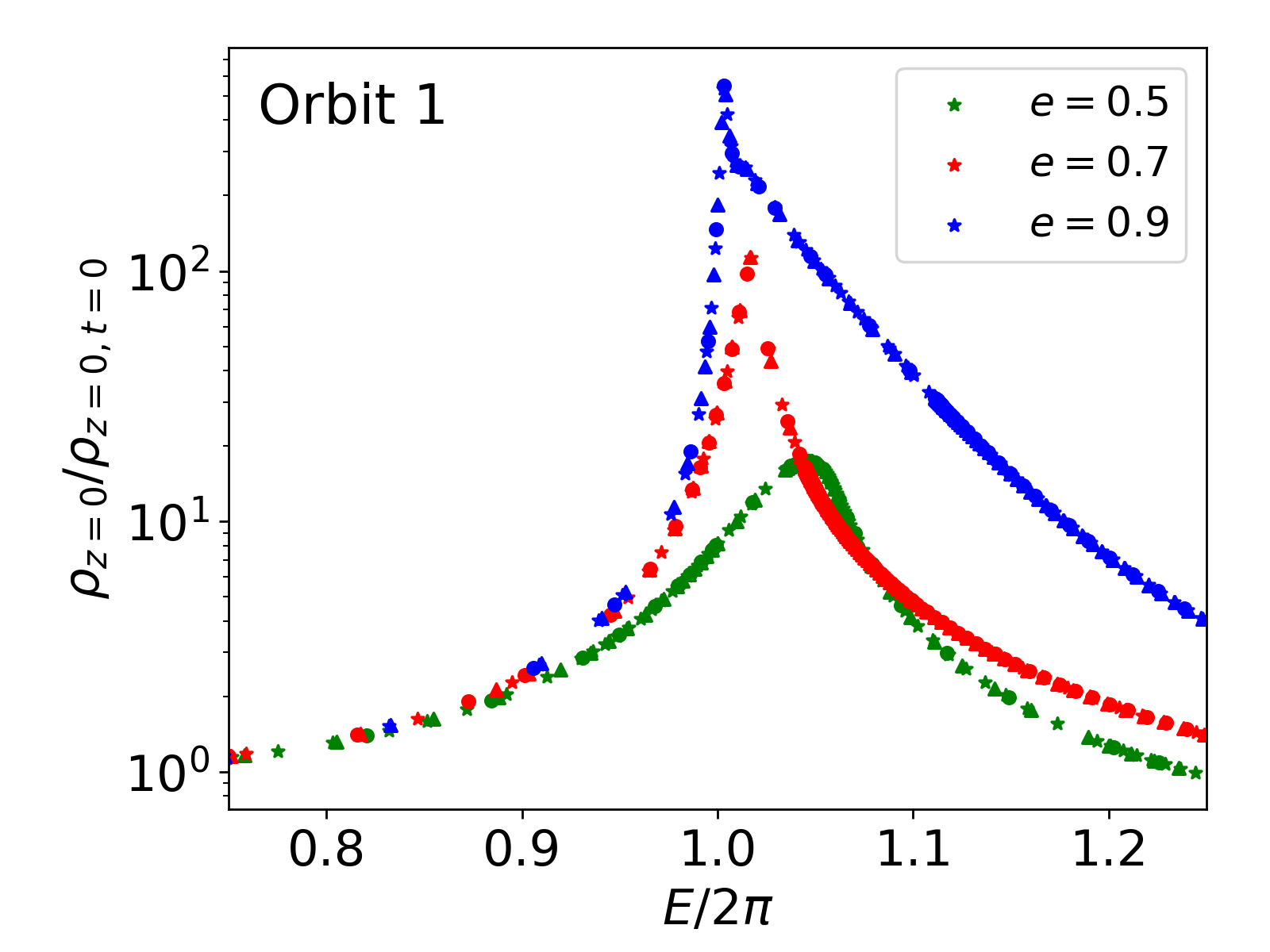}	
\includegraphics[width=8.5cm]{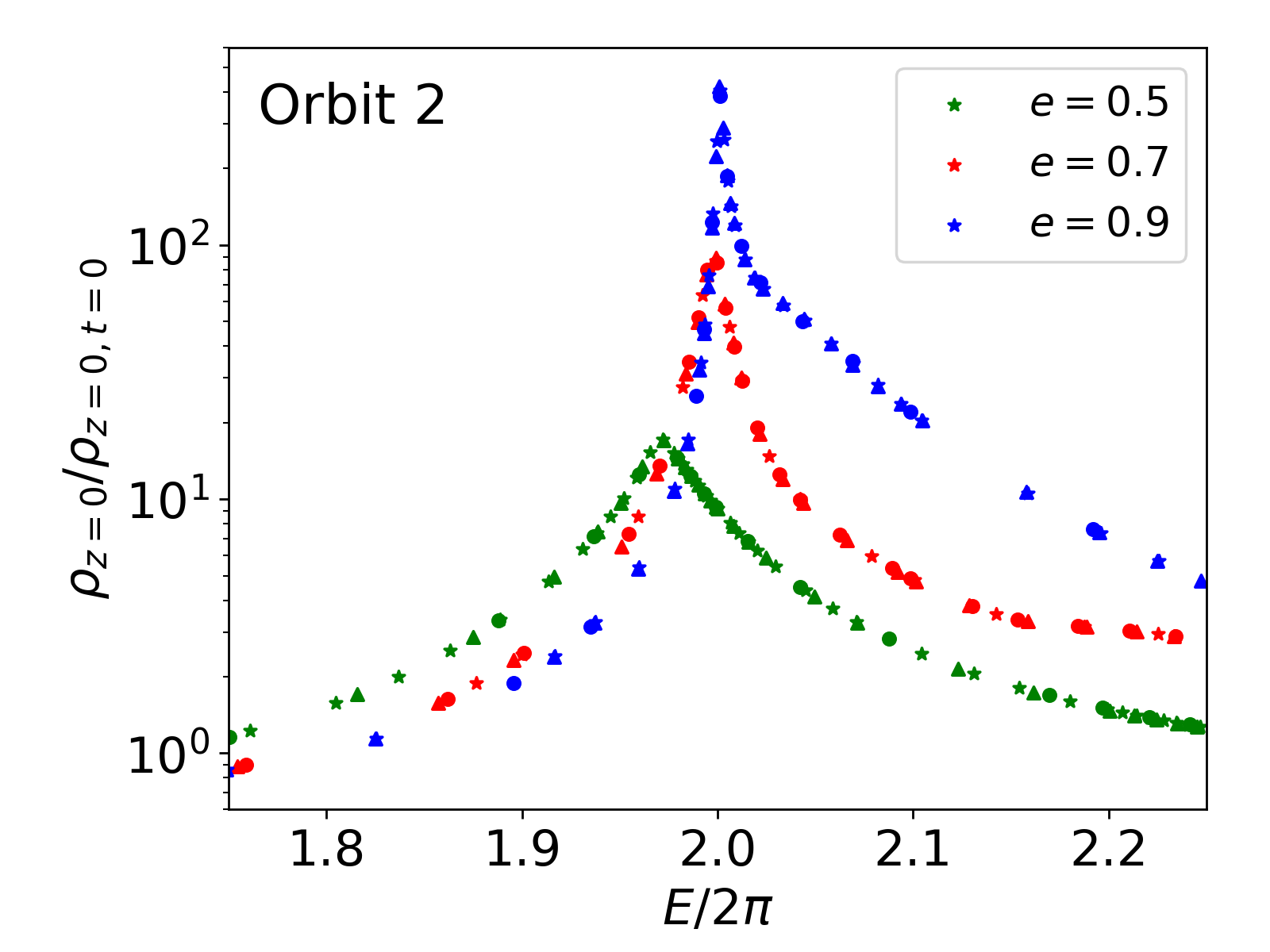}	
		\caption{Converging behaviors of the central density for $e=0.5$, 0.7 and 0.9 (the model \textbf{H01}) with different resolutions near the first (the \textit{left} panel) and second (the \textit{right} panel) pericenter passages. For each $e$, we performed two additional simulations with resolutions that are higher (triangles) and lower (circles) than that for our fiducial cases (circles) by a factor of 1.5. }
	\label{fig:resolution}
\end{figure*}

The disk center has the shortest lengthscale near pericenter where the gas is most compressed. So the converging behavior of the central density can be indicative of the accuracy of our calculations. In Figure~\ref{fig:resolution}, we show the evolution of the central density, normalized with its initial value, $\rho_{z=0}/\rho_{z=0,t=0}$, for $e=0.5$, $0.7$ and $0.9$ near the first (the \textit{left} panel) and second (the \textit{right} panel) pericenter passages.  It is clear that the evolution in the simulations with three different resolutions is nearly identical.

\end{document}